\documentclass[12pt]{iopart}
\usepackage{iopams}  
\usepackage{geometry}
\usepackage{graphicx}
\usepackage{tikz}
\usetikzlibrary{snakes}
\usepackage{stackrel}

\newcommand{\Free}{
\begin{tikzpicture}[scale=0.3]
\draw [black, line width=0.1] (0,0) -- (0,1); 
\draw [black, line width=0.1] (0,1) -- (1,1); 
\draw [black, line width=0.1] (1,1) -- (1,0); 
\draw [black, line width=0.1] (1,0) -- (0,0); 
\end{tikzpicture}}

\newcommand{\Left}{
\begin{tikzpicture}[scale=0.3]
\draw [black, line width=1.5] (0,0) -- (0,1); 
\draw [black, line width=0.1] (0,1) -- (1,1); 
\draw [black, line width=0.1] (1,1) -- (1,0); 
\draw [black, line width=0.1] (1,0) -- (0,0); 
\end{tikzpicture}}

\newcommand{\Low}{
\begin{tikzpicture}[scale=0.3]
\draw [black, line width=0.1] (0,0) -- (0,1); 
\draw [black, line width=0.1] (0,1) -- (1,1); 
\draw [black, line width=0.1] (1,1) -- (1,0); 
\draw [black, line width=1.5] (1,0) -- (0,0); 
\end{tikzpicture}}

\newcommand{\LeftRight}{
\begin{tikzpicture}[scale=0.3]
\draw [black, line width=1.5] (0,0) -- (0,1); 
\draw [black, line width=0.1] (0,1) -- (1,1); 
\draw [black, line width=1.5] (1,1) -- (1,0); 
\draw [black, line width=0.1] (1,0) -- (0,0); 
\end{tikzpicture}}

\newcommand{\LeftLow}{
\begin{tikzpicture}[scale=0.3]
\draw [black, line width=1.5] (0,0) -- (0,1); 
\draw [black, line width=0.1] (0,1) -- (1,1); 
\draw [black, line width=0.1] (1,1) -- (1,0); 
\draw [black, line width=1.5] (1,0) -- (0,0); 
\end{tikzpicture}}

\newcommand{\LowUp}{
\begin{tikzpicture}[scale=0.3]
\draw [black, line width=0.1] (0,0) -- (0,1); 
\draw [black, line width=1.5] (0,1) -- (1,1); 
\draw [black, line width=0.1] (1,1) -- (1,0); 
\draw [black, line width=1.5] (1,0) -- (0,0); 
\end{tikzpicture}}

\newcommand{\LeftLowRight}{
\begin{tikzpicture}[scale=0.3]
\draw [black, line width=1.5] (0,0) -- (0,1); 
\draw [black, line width=0.1] (0,1) -- (1,1); 
\draw [black, line width=1.5] (1,1) -- (1,0); 
\draw [black, line width=1.5] (1,0) -- (0,0); 
\end{tikzpicture}}

\newcommand{\LeftLowUp}{
\begin{tikzpicture}[scale=0.3]
\draw [black, line width=1.5] (0,0) -- (0,1); 
\draw [black, line width=1.5] (0,1) -- (1,1); 
\draw [black, line width=0.1] (1,1) -- (1,0); 
\draw [black, line width=1.5] (1,0) -- (0,0); 
\end{tikzpicture}}

\newcommand{\All}{
\begin{tikzpicture}[scale=0.3]
\draw [black, line width=1.5] (0,0) -- (0,1); 
\draw [black, line width=1.5] (0,1) -- (1,1); 
\draw [black, line width=1.5] (1,1) -- (1,0); 
\draw [black, line width=1.5] (1,0) -- (0,0); 
\end{tikzpicture}}

\newcommand{\freeside}{
\begin{tikzpicture}[scale=0.3]
\draw [black, line width=0.1] (0,0) -- (0,1); 
\end{tikzpicture}}

\newcommand{\bdside}{
\begin{tikzpicture}[scale=0.3]
\draw [black, line width=1.5] (0,0) -- (0,1); 
\end{tikzpicture}}

\newcommand{\freecorn}{
\begin{tikzpicture}[scale=0.3]
\draw [black, line width=0.1] (0,0) -- (0,1); 
\draw [black, line width=0.1] (1,0) -- (0,0); 
\end{tikzpicture}}

\newcommand{\mixtcorn}{
\begin{tikzpicture}[scale=0.3]
\draw [black, line width=1.5] (0,0) -- (0,1); 
\draw [black, line width=0.1] (1,0) -- (0,0); 
\end{tikzpicture}}

\newcommand{\bdcorn}{
\begin{tikzpicture}[scale=0.3]
\draw [black, line width=1.5] (0,0) -- (0,1); 
\draw [black, line width=1.5] (1,0) -- (0,0); 
\end{tikzpicture}}

\newcommand{\triangulartriangular}{
\begin{tikzpicture}[scale=1.5*1.5/1.7]
\draw[black, line width=0.3mm] (4+2/3,2*1.732/3) -- (4+4/3,2*1.732/3);
\draw[black, line width=0.3mm] (4+1/3,1.732/3) -- (4+5/3,1.732/3);
\draw[black, line width=0.3mm] (4,0) -- (6,0);
\draw[black, line width=0.3mm] (5,1.732) -- (4,0); 
\draw[black, line width=0.3mm] (5,1.732) -- (6,0); 
\draw[black, line width=0.3mm] (4+2/3,2*1.732/3) -- (4+4/3,0); 
\draw[black, line width=0.3mm] (4+1/3,1.732/3) -- (4+2/3,0);
\draw[black, line width=0.3mm] (4+2/3,0) -- (4+4/3,2*1.732/3);
\draw[black, line width=0.3mm] (4+4/3,0) -- (4+5/3,1.732/3);
\draw (5,-0.25) node {$M$};
\end{tikzpicture}}

\newcommand{\triangularparallel}{
\begin{tikzpicture}[scale=1.5*1.5/1.7]
\draw[black, line width=0.3mm] (4+2/3,2*1.732/3) -- (4+10/3,2*1.732/3);
\draw[black, line width=0.3mm] (4+1/3,1.732/3) -- (4+9/3,1.732/3);
\draw[black, line width=0.3mm] (4,0) -- (6+2/3,0);
\draw[black, line width=0.3mm] (5,1.732) -- (4,0); 
\draw[black, line width=0.3mm] (5,1.732) -- (6,0); 
\draw[black, line width=0.3mm] (5,1.732) -- (7+2/3,1.732); 
\draw[black, line width=0.3mm] (6+2/3,0) -- (7+2/3,1.732);
\draw[black, line width=0.3mm] (6,0) -- (7,1.732);
\draw[black, line width=0.3mm] (4+4/3,0) -- (4+2/3,2*1.732/3);
\draw[black, line width=0.3mm] (6+1/3,1.732) -- (4+4/3,0); 
\draw[black, line width=0.3mm] (4+1/3,1.732/3) -- (4+2/3,0);
\draw[black, line width=0.3mm] (4+2/3,0) -- (4+5/3,1.732);
\draw[black, line width=0.3mm] (4+4/3,0) -- (4+5/3,1.732/3);
\draw[black, line width=0.3mm] (6+2/3,0) -- (5+2/3,1.732);
\draw[black, line width=0.3mm] (7,1.732/3) -- (6+1/3,1.732);
\draw[black, line width=0.3mm] (7+1/3,2*1.732/3) -- (7,1.732);
\draw (5+1/3,-0.25) node {$M$};
\draw (7+1.5/3,1.7/2) node {$N$};
\end{tikzpicture}}

\newcommand{\squarelattice}{
\begin{tikzpicture}[scale=1.5]
\foreach \i in {0,...,4} 
\draw[black, line width=0.3mm] (0.5*\i,0) -- (0.5*\i,1.5); 
\foreach \i in {0,...,3} 
\draw[black, line width=0.3mm] (0,0.5*\i) -- (2,0.5*\i); 
\draw (1,-0.25) node {$M$}; 
\draw (2.25,0.75) node {$N$};
\end{tikzpicture}}

\begin{document}

\title[Corner free energies and boundary effects]
{Corner free energies and boundary effects for
Ising, Potts and fully-packed loop models on the square and
triangular lattices}

\author{Eric Vernier$^{1,2}$ and Jesper Lykke Jacobsen$^{1,3,4}$}
\address{${}^1$LPTENS, \'Ecole Normale Sup\'erieure, 24 rue Lhomond, 75231 Paris, France}
\address{${}^2$Institut de Physique Th\'eorique, CEA Saclay, 91191
  Gif-sur-Yvette, France}
\address{${}^3$Universit\'e Pierre et Marie Curie, 4 place Jussieu, 75252 Paris, France}
\address{${}^4$Institut Henri Poincar\'e, 11 rue Pierre et Marie Curie,
  75231 Paris, France}

\eads{\mailto{eric.vernier@ens.fr},
      \mailto{jesper.jacobsen@ens.fr}}

\begin{abstract}

  We obtain long series expansions for the bulk, surface and corner
  free energies for several two-dimensional statistical models, by
  combining Enting's finite lattice method (FLM) with exact transfer
  matrix enumerations. The models encompass all integrable curves of
  the $Q$-state Potts model on the square and triangular lattices,
  including the antiferromagnetic transition curves and the Ising
  model ($Q=2$) at temperature $T$, as well as a fully-packed O($n$)
  type loop model on the square lattice. The expansions are around the
  trivial fixed points at infinite $Q$, $n$ or $1/T$.

  By using a carefully chosen expansion parameter, $q \ll 1$, all
  expansions turn out to be of the form $\prod_{k=1}^\infty
  (1-q^k)^{\alpha_k + k \beta_k}$, where the coefficients $\alpha_k$
  and $\beta_k$ are periodic functions of $k$. Thanks to this
  periodicity property we can conjecture the form of the expansions to
  all orders (except in a few cases where the periodicity is too
  large). These expressions are then valid for all $0 \le q < 1$.

  We analyse in detail the $q \to 1^-$ limit in which the models
  become critical. In this limit the divergence of the corner free
  energy defines a universal term which can be compared with the conformal
  field theory (CFT) predictions of Cardy and Peschel. This allows us
  to deduce the asymptotic expressions for the correlation length in
  several cases.

  Finally we work out the FLM formulae for the case where some of the
  system's boundaries are endowed with particular (non-free) boundary
  conditions. We apply this in particular to the square-lattice Potts
  model with Jacobsen-Saleur boundary conditions, conjecturing the
  expansions of the surface and corner free energies to arbitrary
  order for any integer value of the boundary interaction parameter
  $r$. These results are in turn compared with CFT predictions.

\end{abstract}

% \pacs{}
% \submitto{\JPA}

\section{Introduction}

Over the years much effort has been devoted to the study of boundary
effects in statistical models. In the particular case of critical
two-dimensional systems a huge amount of knowledge has been obtained
by the application of the powerful techniques of integrability and
conformal field theory (CFT) \cite{Dosch,Brankov}. The critical
fluctuations near a boundary have been shown to define various
boundary critical exponents, many of which can be computed exactly.

Once a model has been shown to be exactly solvable (integrable), it is
usually rather simple to obtain the dominant asymptotics for the
partition function, i.e., the bulk free energy $f_{\rm b}$. The usual
route is to diagonalise the transfer matrix in a cylinder geometry,
using the Bethe Ansatz, and to extract its dominant eigenvalue in the
thermodynamic limit. Similarly, if the boundary conditions are
compatible with integrability (in the sense of Sklyanin's reflection
equation \cite{Sklyanin}) one can diagonalise the transfer matrix in a
strip geometry and retrieve the surface free energy $f_{\rm s}$.
Obviously $f_{\rm b}$ and $f_{\rm s}$ give important information about
the particular lattice model being studied. However, both quantities
are non-universal and as such teach us nothing about the critical
system that emerges in the continuum limit.

If the lattice model under study is defined on a large rectangle, the
next subdominant contribution to the free energy comes from the
corners. Cardy and Peschel \cite{CardyPeschel} have shown that this
corner free energy $f_{\rm c}$ is of universal nature. Indeed it
contains information about the central charge $c$ of the field theory
that governs the continuum limit.

Unfortunately only very little is known about $f_{\rm c}$ from the
point of view of exactly solvable models. To compute it from the Bethe
Ansatz would require knowledge about how to implement a given boundary
condition in terms of Bethe states. Typically this would call for
information about all the eigenstates, not just the one defining the
leading eigenvalue. This obviously defines a very hard problem which,
as far as we know, has not yet been addressed.

One of the authors has recently conjectured exact product formulae for
${\rm e}^{f_{\rm b}}$, ${\rm e}^{f_{\rm s}}$ and ${\rm e}^{f_{\rm c}}$
in a particular two-dimensional model, viz.\ the zero-temperature
antiferromagnetic $Q$-state Potts model on the triangular lattice
\cite{JacobsenChromatic}. To obtain these expressions, the first step
was to compute the first 40 terms of their large-$Q$ expansion, by
combining Enting's finite lattice method (FLM) \cite{Enting1} with
exact transfer matrix enumerations. The next step was to notice that
when reexpressed in terms of the variable $x \ll 1$ defined by $Q = 2
- x - x^{-1}$ these series could be rewritten as product formulae of
the type
\begin{equation}
 \prod_{k=1}^\infty (1-x^k)^{\alpha_k} \,,
\end{equation}
where crucially the exponents $\alpha_k$ turned out to be {\em
  periodic} functions of $k$. Since the observed periods (6 for
$f_{\rm b}$, and 12 for $f_{\rm s}$ and $f_{\rm c}$) were much shorter
than the number of available terms, it then became feasible to
conjecture the exact product formulae, valid for any $x$ in the range
$0 \le x < 1$.  The model is non-critical in that range, but goes to a
critical theory when $x \to 1^-$ (i.e., $Q \to 4^+$).

The purpose of this paper is to extend this type of results to a whole
range of two-dimensional lattice models. We recall the definitions of
the models to be studied in section~\ref{sec:models} below.  They
encompass all integrable curves of the $Q$-state Potts model on the
square and triangular lattices, including the antiferromagnetic
transition curves and the Ising model ($Q=2$) at temperature $T$, as
well as a fully-packed O($n$) type loop model on the square
lattice. The expansions are around the trivial fixed points at
infinite $Q$, $n$ or $1/T$.

A crucial ingredient in carrying out this programme is obviously to
identify the correct expansion variable $x$ (that we shall call $q$ in
the general case). We wish $q$ to have the property that the model is
trivial for $q=0$, non-critical for $0 \le q < 1$, and critical in the
limit $q \to 1^-$. Below we shall focus only on models which are known
to be integrable (at least in the bulk), and we can therefore take
inspiration for the choice of $q$ from the exact solution (for the
bulk properties). Incidentally, we have no reason to believe that
nice product formulae hold for models that are not integrable.

In all cases that we have investigated, it appears that
${\rm e}^{f_{\rm b}}$, ${\rm e}^{f_{\rm s}}$ and ${\rm e}^{f_{\rm c}}$
can be cast as exact product formulae of the more general form
\begin{equation}
 \prod_{k=1}^\infty (1-q^k)^{\alpha_k} \,, \qquad
 \mbox{with } \alpha_k = \beta_k k + \gamma_k \,,
 \label{generic factorized form}
\end{equation}
where both sets of coefficients $\beta_k$ and $\gamma_k$ are periodic
functions of $k$ with the same period. We find for all Potts and
fully-packed loop models included in this study that the period
associated with ${\rm e}^{f_{\rm s}}$ and ${\rm e}^{f_{\rm c}}$ is
precisely twice the period associated with ${\rm e}^{f_{\rm b}}$.  For
Ising models we find on the contrary that all three quantities have
the same period. At present we have no satisfactory explanation for
this observation of period doubling for the boundary related
properties.

Just like in \cite{JacobsenChromatic} we can then conjecture the exact
product formula, provided that the series obtained from the FLM has
sufficiently many terms to cover at least one period (plus a few extra
terms to verify the assumption of periodicity, and to account for some
simple extra factors in front of the product which are sometimes
present). Only in a few exceptional cases do the series turn out to be
too short, but we can then at least state how many further terms would
be needed in order to establish the product form.

The organisation of the paper is as follows. In
section~\ref{sec:models} we precisely define the models to be studied
in this paper. In section~\ref{sec:free_bcs} we first review the FLM
for models with free boundary conditions. Then we present our main
results in the form of product formulae for the bulk, surface and corner
free energies of the various models. We give details about the period
doubling phenomenon for the boundary related properties. When studying
the critical limit $q \to 1^-$ we pay special attention to the
divergence of the corner free energy, since this can be compared with
CFT results \cite{CardyPeschel}. As a by-product we obtain results
about the asymptotic behaviour of the correlation length.

In section~\ref{sec:boundaries} we study the effect of imposing
particular (non-free) boundary conditions on some of the system's
boundaries. We adapt the FLM formalism to this case and separate the
contributions from corners of different types, i.e., where two free
(resp.\ two particular, resp.\ one free and one particular) boundary
conditions meet. It is shown analytically that the contributions to
the free energy from corners of different types simply add up, as
expected for an inherently non-critical system. We apply this
formalism to a family of Potts-model boundary conditions recently
introduced by Jacobsen and Saleur (JS)
\cite{JacobsenSaleurConformalBoundary} in which a parameter $r$
controls the weight of Fortuin-Kasteleyn clusters that touch the
particularised boundaries. Explicit results are obtained for any
integer value of $r$.  To finish, we comment on the relation with CFT
results for such boundary conditions.

\section{Models}
\label{sec:models}

In this section we define the models to be studied in this paper and
briefly review their most relevant properties. These models constitute
all integrable cases of the Potts model on the square and triangular
lattices. We also study a model of two mutually excluding sets of
fully-packed loops, known as the ${\rm FPL}^2$ model
\cite{KondevHenley}.

In the entire paper we will define the (dimensionless) free energy as
the logarithm of the partition function, $f \equiv \ln Z$, i.e., {\em
  without} the conventional minus sign.

\subsection{Potts model: generalities}

Let us recall that the $Q$-state Potts model is a model of interacting
spins on a lattice, allowing each spin to be in one among $Q$
different states, and such that the interaction between neighbouring
spins depends only on whether they are in the same or different
states. The underlying lattice, or graph, is denoted as $G=(V,E)$,
where $V$ (resp.\ $E$) is the set of vertices (resp.\ edges).  The
associated (dimensionless) hamiltonian thus reads
\begin{equation}
 \beta H =
 -K \sum_{(i,j) \in E}
 \delta_{\sigma_i, \sigma_j} \,, \qquad \sigma_i = 1,\ldots,Q \,,
 \label{H_Potts}
\end{equation}
where $K=J/k_B T$ is the (dimensionless) coupling constant.

It is well-known that the partition function can be rewritten as \cite{FK}
\begin{equation}
 Z = \sum_{A \subseteq E} Q^{k(A)} v^{|A|} \,,
 \label{Z_Potts_FK}
\end{equation}
where the sum runs over subsets $A$ (of cardinality $|A|$) of the
lattice edges $E$. Each connected component in $A$ is known as a
Fortuin-Kasteleyn (FK) cluster, and $k(A)$ denotes the number of
connected components. The temperature parameter $v$ is defined as $v =
{\rm e}^K-1$.

Obviously, the FK formulation makes sense also when $Q$ is an
arbitrary real number. Suppose now that the Potts model is defined on
an infinite regular lattice ($|V|,|E| \to \infty$) with coordination
number $z=2|E|/|V|$. If we study the model on some curve in the
$(Q,v)$ plane with asymptotics
\begin{equation}
 v \sim Q^{2/z} \qquad \mbox{for } |Q| \gg 1 \,,
 \label{phasecoexistence}
\end{equation}
then there will be precisely two dominant contributions to $Z$: $A =
E$ (with $k(A) = 1$) and $A = \emptyset$ (with $k(A) = |V|$). In other
words, there is phase coexistence between the low-temperature and
high-temperature phases. All subdominant contributions can be written
perturbatively as powers of the small parameter $1/Q$. This perturbative picture
is the starting point for making series expansions of the free energy.

\subsection{Lattices and their orientation}

In a detailed study of boundary effects it is important to specify not
only on which regular lattice the model is defined, but also how the
boundaries are oriented with respect to the lattice's symmetry axes.

Moreover, we do not expect that the free energy series can be written
as nice product formulae of the type (\ref{generic factorized form}) for
an arbitrary lattice model of interest. The model will in general have
to be {\em integrable} to yield nice expressions. We recall that a
model remains integrable in the presence of boundaries, only if the
boundary conditions satisfy the reflection equation \cite{Sklyanin}.
This places strong constraints on how the boundaries can be oriented.

\begin{figure}
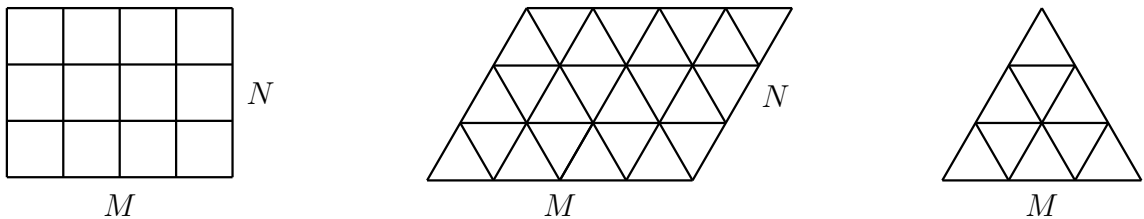

\begin{center}
\squarelattice \qquad \qquad \triangularparallel \qquad \qquad \triangulartriangular
\caption{Different types of lattices to be used throughout this
  work. From left to right: a) square lattice in a rectangle of size
  $M\times N$, b) triangular lattice in a (deformed) rectangle of size $M
  \times N$, and c) triangular lattice in an equilateral triangle of size
  $M$.}
\label{lattices}
\end{center}
\end{figure}

The Potts model on the square and triangular lattices is integrable
for certain curves in the $(Q,v)$ plane, which we briefly review
below.  Boundary integrability turns out to hold as well, provided
that the boundaries are oriented parallel to the principal axes of the
lattice. In particular, an $M \times N$ rectangular piece of the
square lattice will be oriented as shown in Fig.~\ref{lattices}a.

When dealing with a triangular lattice, we can again take an $M \times
N$ (deformed) rectangular piece, as shown in Fig.~\ref{lattices}b.
This lattice can be considered simply as a square lattice with added
diagonals. This point of view is often convenient, since then the FLM
formulae can be taken over from the square-lattice case, and the
transfer matrix algorithms need only very minor modifications.

There is however one disadvantage: since the four corners are not
equivalent, the corner free energy will be a mixture of two
contributions from corners sustaining an angle $\pi/3$ and two
contributions from $2\pi/3$ corners. To separate these contributions
we shall find it advantageous to consider as well the triangular
lattice inscribed in an equilateral triangle of side length $M$, as
shown in Fig.~\ref{lattices}c. This involves three $\pi/3$ corners,
but makes both the FLM formulae and the transfer matrix algorithms
slightly less performing.

\subsection{Square-lattice Potts model}
\label{sec:sq_latt_potts}

\begin{figure}
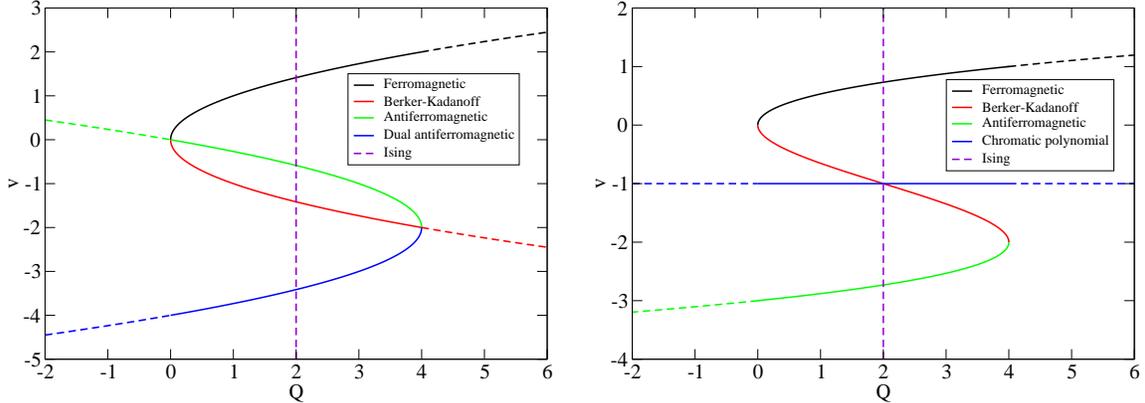

\begin{center}
\includegraphics[width=0.48\textwidth]{square_pd.eps} \quad
\includegraphics[width=0.48\textwidth]{triangular_pd.eps}
\caption{Phase diagrams in the $(Q,v)$ plane of the Potts model on a)
  the square lattice, and b) the triangular lattice. The models are
  integrable along the curves shown. Critical (resp.\ non-critical)
  behaviour is signalled by solid (resp.\ dashed) line style. For
  further details on these curves, their properties and nomenclature,
  and the renormalisation group flows, the reader can consult
  \cite{Saleur91,JacobsenSaleurAntiferro} and references therein.}
\label{phasediagrams}
\end{center}
\end{figure}

The phase diagram of the square-lattice Potts model is shown in
Fig.~\ref{phasediagrams}a. The model is integrable \cite{squareBaxter}
along the selfdual curves
\begin{equation}
 v = \pm \sqrt{Q} \,,
 \label{squarePottscritical}
\end{equation}
corresponding to a critical (resp.\ non-critical) theory for $0 \le Q
\le 4$ (resp.\ $Q > 4$). 
% The model is invariant under a simultaneous
% sign change of $v$ and $\sqrt{Q}$, so the curve $v = -\sqrt{Q}$ can be
% attained as well by taking $v > 0$ and the other branch of the square
% root.
The conformal properties in the critical regime are known
\cite{AlcarazBarberBatchelor}.  The only simple property that we need
here is that the point $(Q,v) = (4,2)$ is critical with central charge
$c=1$.

Note that (\ref{squarePottscritical}) satisfies the criterion
(\ref{phasecoexistence}), thus making possible a perturbative expansion for
$Q \gg 1$. We shall see below that this expansion enables us to
attain the point $(Q,v) = (4,2)$ as an appropriate limit.

The square-lattice Potts model is also integrable \cite{AFBaxter}
along the mutually dual antiferromagnetic curves
\begin{equation}
 v = -2 \pm \sqrt{4-Q} \,,
 \label{AFPottscritical}
\end{equation}
corresponding to a critical (resp.\ non-critical) theory for $0 \le Q
\le 4$ (resp.\ $Q < 0$).

The conformal properties along (\ref{AFPottscritical}) are quite
intricate, but much is known
\cite{Saleur91,JacobsenSaleurAntiferro,IkhlefJacobsenSaleur,BlackHole}. In
particular one has $c=-1$ in the limit $(Q,v) \to (0,0)$. Since
(\ref{AFPottscritical}) again satisfies (\ref{phasecoexistence}) the
perturbative expansion enables us to attain this point as an
appropriate limit.

Note finally that the $Q$-state Potts model is equivalent to a problem
of self-avoiding loops \cite{BaxterKellandWu}. The loops are defined
on the medial lattice and each one carries a weight $n = \sqrt{Q}$.
The loop formulation can further be brought in equivalence with the
six-vertex model \cite{BaxterKellandWu}. The six-vertex model is
homogeneous on the curve (\ref{squarePottscritical}) and staggered on
(\ref{AFPottscritical}) \cite{BaxterBook}.

\subsection{Triangular-lattice Potts model}
\label{sec:triangular-lattice_Potts}
The phase diagram of the triangular-lattice Potts model is shown in
Fig.~\ref{phasediagrams}b. The model is integrable along the cubic
curve \cite{triangularBaxter}
\begin{equation}
 v^3 + 3v^2 = Q \,.
 \label{triangularPottscritical}
\end{equation}
Once again (\ref{phasecoexistence}) is satisfied.

For $0 \le Q \le 4$ the critical behaviour for $v \ge 0$ (i.e., along
the upper branch of (\ref{triangularPottscritical})) coincides with
that of the square lattice. Along the lower branch of
(\ref{triangularPottscritical}) one finds
\cite{JacobsenSaleurAntiferro} the same critical behaviour as on the
antiferromagnetic curve (\ref{AFPottscritical}) on the square lattice.

The two critical points that can be accessed perturbatively are $(Q,v)
= (4,1)$ with $c=1$, and $(Q,v) \to (0,-2)$ with $c=-1$
\cite{JacobsenSaleurAntiferro}.

We shall sometimes refer to (\ref{triangularPottscritical}) as the
{\em selfdual curve}. Indeed, if one combines a duality transformation
of the triangular-lattice Potts model with a partial summation (decimation)
over one half of the spins in the dual model on the hexagonal lattice,
the result is a non-trivial transformation whose fixed points are exactly
(\ref{triangularPottscritical}). In this sense the nomenclature ``selfdual''
is justified when speaking about the curve (\ref{triangularPottscritical}).

In the special case $v=-1$ and any real $Q$, the Potts partition
function $Z$ (\ref{Z_Potts_FK}) reduces to the so-called chromatic
polynomial. For integer $Q$ this can be interpreted as a colouring
problem. To be precise, $Z$ equals the number of proper vertex
colourings of the lattice, i.e., assignations to each of the vertices
in $G$ of one among the $Q$ different colours, in such a way that
adjacent vertices carry different colours.

The chromatic polynomial is integrable \cite{Baxter86,Baxter87} for any
real $Q$. It corresponds to a critical theory if and only if $0 \le Q \le 4$.
Its phase diagram within the critical phase is complicated \cite{Baxter87}
and has been reviewed in \cite{JacobsenSalasSokal}.  The only properties
that we shall need here are that the point $(Q,v) = (4,-1)$ has $c=2$,
while the limit $(Q,v) \to (0,-1)$ corresponds to $c=-1$.

The bulk, surface and corner free energies for the chromatic
polynomial on the triangular lattice have already been discussed in
\cite{JacobsenChromatic}. In section~\ref{sec:chrom_poly_tri} we shall
however be able to go beyond these results and separate the
contributions to the corner free energy for two different types of
corners (of angles $\frac{2\pi}{3}$ and $\frac{\pi}{3}$).

\subsection{Ising model}

For $Q=2$ the Potts model reduces to the Ising model with
(dimensionless) hamiltonian
\begin{equation}
  \beta H = -K_{\rm Ising} \sum_{(i,j) \in E} S_i S_j \,,
  \qquad S_i = \pm 1 \,.
 \label{H_Ising}
\end{equation}
The coupling constants in (\ref{H_Potts}) and (\ref{H_Ising}) are
obviously related by $K_{\rm Ising} = \frac12 K$.

The Ising model is integrable at any temperature on both the square
and the triangular lattices \cite{BaxterBook}. We shall only be
interested in the cases $K_{\rm Ising} > 0$. The critical point on the
ferromagnetic critical curves is situated at $K_{\rm Ising} = \frac12
\log(1+\sqrt{2})$ on the square lattice (see
Fig.~\ref{phasediagrams}a) and at $K_{\rm Ising} = \frac14 \log 3$ on
the triangular lattice (see Fig.~\ref{phasediagrams}b). The
low-temperature perturbative expansion gives access to the critical
point in an appropriate limit.

When studying the Ising model, we will always treat it as a Potts
model and rewrite $K_{\rm Ising}$ in terms of $K$.

\subsection{${\rm FPL}^2$ model}

\begin{figure}
\begin{center}
\begin{tikzpicture}[scale=1.0]
  \draw[red,line width=1mm] (0,0.5)--(1,0.5);
  \draw[black,line width=1mm] (0.5,0)--(0.5,1);
  \draw[red,line width=1mm] (2.5,0)--(2.5,1);
  \draw[black,line width=1mm] (2,0.5)--(3,0.5);
  \draw[black,line width=1mm] (4,0.5)--(4.2,0.5) arc(270:360:3mm) -- (4.5,1);
  \draw[red,line width=1mm] (4.5,0)--(4.5,0.2) arc(180:90:3mm) -- (5,0.5);
  \draw[red,line width=1mm] (6,0.5)--(6.2,0.5) arc(270:360:3mm) -- (6.5,1);
  \draw[black,line width=1mm] (6.5,0)--(6.5,0.2) arc(180:90:3mm) -- (7,0.5);
  \draw[black,line width=1mm] (8,0.5)--(8.2,0.5) arc(90:0:3mm) -- (8.5,0);
  \draw[red,line width=1mm] (8.5,1)--(8.5,0.8) arc(180:270:3mm) -- (9,0.5);
  \draw[red,line width=1mm] (10,0.5)--(10.2,0.5) arc(90:0:3mm) -- (10.5,0);
  \draw[black,line width=1mm] (10.5,1)--(10.5,0.8) arc(180:270:3mm) -- (11,0.5);
\end{tikzpicture}
\end{center}
\caption{Allowed vertex states in the FPL${}^2$ model.}
\label{fig:FPL2_vertices}
\end{figure}
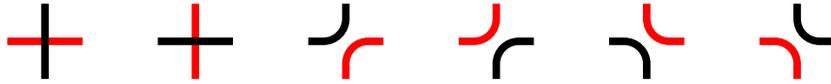

Apart from the special cases of Potts models described above, we also
consider a model of two mutually excluding sets of fully-packed loops
on the square lattice, known as the ${\rm FPL}^2$ model
\cite{KondevHenley}.  The allowed configurations are such that each
vertex is in one of the six states shown in
Fig.~\ref{fig:FPL2_vertices}. A weight $n_1$ (resp.\ $n_2$) is associated
with each black (resp.\ red) loop.

The ${\rm FPL}^2$ model is integrable when the two weights are equal,
$n_1 = n_2 \equiv n$ \cite{DeiCont1,JacobsenZinn,DeiCont2}. We study
henceforth only this integrable case. It gives rise to a critical
(resp.\ non-critical) model when $-2 \le n \le 2$ (resp.\ $|n| > 2$).

When the FPL${}^2$ model is defined on a bipartite 4-regular lattice,
its configurations are in bijection with those of a 3-dimensional
interface model on the dual lattice \cite{KondevHenley}. This allows
for the redistribution of the loop weights $n_1$ and $n_2$ as local
complex vertex weights. In the critical regime the resulting local
formulation is the key to the exact derivation of the critical
exponents by the Coulomb gas method \cite{Kondev,JacobsenKondev}.

\begin{figure}
\begin{center}
\begin{tikzpicture}[scale=0.6]
 \draw[step=10mm,black,line width=0.5mm] (0,0) grid (5,3);
 \draw[black,line width=0.5mm] (0,0)--(0,-0.2) arc(180:360:5mm) -- (1,0);
 \draw[black,line width=0.5mm] (2,0)--(2,-0.2) arc(180:360:5mm) -- (3,0);
 \draw[black,line width=0.5mm] (4,0)--(4,-0.2) arc(180:360:5mm) -- (5,0);
 \draw[black,line width=0.5mm] (0,3)--(0,3.2) arc(180:0:5mm) -- (1,3);
 \draw[black,line width=0.5mm] (2,3)--(2,3.2) arc(180:0:5mm) -- (3,3);
 \draw[black,line width=0.5mm] (4,3)--(4,3.2) arc(180:0:5mm) -- (5,3);
 \draw[black,line width=0.5mm] (0,0)--(-0.2,0) arc(270:90:5mm) -- (0,1);
 \draw[black,line width=0.5mm] (0,2)--(-0.2,2) arc(270:90:5mm) -- (0,3);
 \draw[black,line width=0.5mm] (5,0)--(5.2,0) arc(-90:90:5mm) -- (5,1);
 \draw[black,line width=0.5mm] (5,2)--(5.2,2) arc(-90:90:5mm) -- (5,3);
 \draw[snake=brace] (0,4)--(5,4) node[above=0.4cm,left=1.25cm] {$M$};
 \draw[snake=brace] (6,3)--(6,0) node[right=0.4cm,above=0.65cm] {$N$};
\end{tikzpicture}
\end{center}
\caption{A finite bipartite 4-regular lattice of width $M=3$ and height $N=2$.}
\label{fig:FPL2_lattice}
\end{figure}
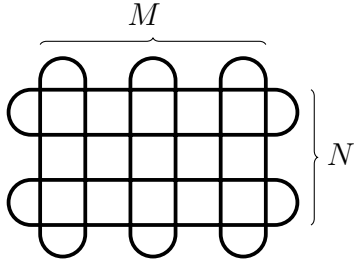

To investigate the surface properties of the FPL${}^2$ model without
breaking any symmetries (which might lead to a change of universality
class) one therefore needs to define it on a finite lattice which is
bipartite and 4-regular \cite{loop_review}. An appropriate choice of
lattice is shown in Fig.~\ref{fig:FPL2_lattice}.

\section{Bulk, surface and corner free energies
 with free boundary conditions}
\label{sec:free_bcs}

In this section, we compute the bulk, surface and corner free energies
of each of the considered models in the case of lattices with free
boundary conditions. Achieving the goals described in the
Introduction relies in all cases on writing these free energies as
power series in terms of some good perturbative parameter. Long series
are then generated by combining transfer matrix techniques with the
use of Enting's finite lattice method \cite{Enting1}, which we first
review below.

For each model being considered, the first term of the series
expansion is some unique (or sometimes twofold degenerate) ground
state. As mentioned at the end of section~\ref{sec:sq_latt_potts} we
shall treat the Potts models (\ref{Z_Potts_FK}) in terms of the
equivalent loop models on the medial lattice. The ground state is then
chosen as the unique state in which one loop surrounds each vertex in
the original lattice $G$; see Fig.~\ref{ground_state}. Equivalently,
the edge subset $A \subseteq E$ appearing in (\ref{Z_Potts_FK}) is
empty, $A = \emptyset$.

Similarly, for the Ising model the ground state is chosen as one of
the two equivalent ferromagnetic states with all spins pointing in the
same direction.

\newcommand{\fundamental}{\begin{tikzpicture}[scale=0.6]
\foreach \x in {0,...,6} 
\foreach \y in {0,...,6} 
   \filldraw [black] (\x,\y) circle (2pt);
\foreach \i in {0,...,6} 
\foreach \j in {0,...,6} 
\draw[red,line width=0.4mm, rounded corners=4pt] (\i+0.5,\j+0) -- (\i+0,\j+0.5
) -- (\i-0.5,\j+0) -- (\i+0,\j-0.5) -- cycle;
\end{tikzpicture}}

\newcommand{\perturbed}{\begin{tikzpicture}[scale=0.6]

\foreach \x in {0,...,6} 
\foreach \y in {0,...,6} 
   \filldraw [black] (\x,\y) circle (2pt);

\foreach \i in {0,...,3} 
\foreach \j in {0,...,6} 
\draw[red,line width=0.4mm, rounded corners=4pt] (\i+0.5,\j+0) -- (\i+0,\j+0.5
) -- (\i-0.5,\j+0) -- (\i+0,\j-0.5) -- cycle;
\foreach \j in {0,...,6} 
\draw[red,line width=0.4mm, rounded corners=4pt] (6+0.5,\j+0) -- (6+0,\j+0.5
) -- (6-0.5,\j+0) -- (6+0,\j-0.5) -- cycle;
\foreach \i in {4,5} 
\foreach \j in {0,4,5,6} 
\draw[red,line width=0.4mm, rounded corners=4pt] (\i+0.5,\j+0) -- (\i+0,\j+0.5
) -- (\i-0.5,\j+0) -- (\i+0,\j-0.5) -- cycle;
\foreach \j in {2,3} 
\draw[red,line width=0.4mm, rounded corners=4pt] (5+0.5,\j+0) -- (5+0,\j+0.5
) -- (5-0.5,\j+0) -- (5+0,\j-0.5) -- cycle;

\draw[black, line width=0.4mm] (5,1) -- (4,1) -- (4,3);
\draw[red, line width=0.4mm, rounded corners=4pt] (5.5,1) -- (5,1.5) -- (4.5,1) -- (4,1.5) -- (4.5,2) -- (4,2.5) -- (4.5,3) -- (4,3.5) -- (3.5,3) -- (4,2.5) -- (3.5,2) -- (4,1.5) -- (3.5,1) -- (4,0.5) -- (4.5,1) -- (5,0.5) -- cycle;
\draw[blue, line width=0.4mm] (5.5,0.5) -- (5.5,3.5)-- (3.5,3.5)-- (3.5,0.5) -- (3.5,0.5) -- cycle;

\end{tikzpicture}}

\begin{figure}
\begin{center}
\fundamental \qquad \qquad \perturbed
\end{center}
\caption{Representation of the ground state used as a basis for the
  finite lattice method (left), and of one of the first excitations to
  be taken into account in the series expansion (right), in the case of
  the square-lattice Potts model. The black edges on the right
  correspond to the subset $A \subseteq E$ appearing in
  (\ref{Z_Potts_FK}). The blue box surrounding $A$ corresponds to the
  smallest finite sublattice $[i,j]$ in which this excitation
  appears.}
\label{ground_state}
\end{figure}

For the ferromagnetic Potts model (both on the square and
triangular lattices), the chromatic polynomial on the triangular
lattice, or for the ${\rm FPL}^2$ model, a good expansion parameter will
turn out to be, rather than the inverse $n^{-1}$ of the loop fugacity ($n\equiv \sqrt{Q}$),
the parameter $q$
defined by
\begin{equation}
 \sqrt{Q} = q + \frac{1}{q} \,.
\end{equation}
Actually, for these models $q$ is nothing but the quantum group
deformation parameter. Indeed, all these Potts models present a quantum
group symmetry $U_q(sl_2)$. The ${\rm FPL}^2$ model is endowed with
the larger symmetry $U_q(sl_4)$.

For the critical antiferromagnetic Potts model on the square lattice
the suitable parameter $q$ is defined by
\begin{equation}
 Q=-\left(q - \frac{1}{q}\right)^2 \,.
\end{equation}

Finally, for the Ising model the choice of the expansion parameter $q$
is far less obvious. Indeed, the approach outlined in the Introduction
makes it necessary that the critical point correspond to the limit $q
\to 1^-$. The correct choice of $q$ then involves parameterising the
coupling constant in terms of suitable elliptic functions, as in
Baxter's analytical results \cite{BaxterIsing,BaxterBook}. We defer
further details to the corresponding sections below.

\subsection{FLM for lattices with free boundary conditions}

We now outline the principles of the FLM for
lattices with free boundary conditions. The formalism depends on the
choice of regular polygons in which the excitations over the ground
state are inscribed. The most common and well-known choice is
rectangles, as illustrated by the blue box in Fig.~\ref{ground_state}.
We review this first (in section~\ref{sec:FLM_rectangle}) for
convenience, and in order to fix the notation. Then, in
section~\ref{sec:FLM_triangle}, we derive the required modifications
for the case of triangular lattices with excitations inscribed in
equilateral triangles, cf.\ Fig.~\ref{lattices}c.

\subsubsection{Lattices inscribed in a rectangle.}
\label{sec:FLM_rectangle}

We here give a brief review of the earlier work by Enting
\cite{Enting1} concerning the basis of the finite lattice method.
The FLM allows one to compute the free energies of statistical
models defined on an infinite rectangular lattice as a series
expansion in terms of finite rectangular graphs.  Apart from the
square lattice, this method can also be applied to the triangular
lattice, which is then considered as a square lattice with added
diagonals; see Fig.~\ref{lattices}b.

In practice, we will make extensive use of the FLM to approximate the
bulk, surface and corner free energies of an infinite $M \times N$
lattice, such as depicted in Fig.~\ref{lattices}a--b, from free
energies for finite $m \times n$ lattices. The latter are computed
by using numerical (but exact) transfer matrix enumerations.
The bigger the $m \times n$ lattices whose partition function can be
computed numerically, the longer will be the resulting FLM series.

The basic idea underlying the FLM is to write the free energy for the
$M\times N$ lattice as a sum over all possible sublattices,
\begin{equation}
 f_{M,N} = \sum_{[i,j]\subset[M,N]}{\tilde{f}_{i,j}} 
 = \sum_{i\leq M, j\leq N }{(M-i+1)(N-j+1)\tilde{f}_{i,j}} \,,
\label{fMNexact}
\end{equation}
where the notation $[i,j]\subset[M,N]$ means that the summation is
performed over all sublattices of size $i\times j$ that lie inside the
$M\times N$ lattice. The contributions $\tilde{f}_{i,j}$---that must
not be confused with the $[i,j]$ lattices' free energies
$f_{i,j}$---are defined self-consistently as functions of the latter
by inverting the equation
\begin{equation}
 f_{m,n} = \sum_{[i,j]\subset[m,n]}{\tilde{f}_{i,j}}
 = \sum_{i\leq m, j\leq n}{(m-i+1)(n-j+1)\tilde{f}_{i,j}} \,.
\end{equation}
This results in
\begin{equation}
 \tilde{f}_{i,j} = \sum_{m\leq i, n\leq j}{f_{m,n}\eta(m,i)\eta(n,js)} \,,
 \label{reciprocalftilde}
\end{equation}
where the functions $\eta$ are defined by Enting \cite{Enting1} as
\begin{equation}
 \eta(m,i)=\left\{
 \begin{array}{ll}
	1 & \mbox{ if } m=i  \mbox{ or } m+2=i \mbox{ and } i>2 \,, \\
	2 & \mbox{ if } m+1=i \mbox{ and } i>1 \,, \\
	0 & \mbox{ otherwise.} 
 \end{array}
 \right.
\end{equation}

For an infinite $M \times N$ lattice, the sum (\ref{fMNexact}) is
restricted to a certain range of finite $m \times n$ sublattices by
imposing some cutoff set. In the course of this paper, and as
prescribed in \cite{Enting1}, this cutoff set is chosen to be
\begin{equation}
 B(k) = \left\{ [m,n] \,, m+n=k \right\} \,.
\end{equation}

Hence, (\ref{fMNexact}) formally reads
\begin{equation}
 f_{M,N}\approx \sum_{[i,j]\leq B(k)}{(M-i+1)(N-j+1)\tilde{f}_{i,j}} \,,
 \label{fMNapprox}
\end{equation}
which is generally shown to give an approximation of $f_{M,N}$ up to
an order increasing with $k$ in terms of some good perturbative parameter, as
discussed above and detailed in the following sections.

The free energy (\ref{fMNapprox}) can be written as the sum of a bulk
contribution $MN f_{\rm b}$, a surface contribution $(M+N)f_{\rm s}$,
and a corner contribution $f_{\rm c}$, that is
\begin{equation}
 f_{M,N} = M N f_{\rm b} + (M+N)f_{\rm s} + f_{\rm c}
\label{fMN_decomposition}
\end{equation}
These three contributions can be computed separately.
The final form of the FLM expansions with cutoff $B(k)$ come out as
\cite{Enting1}
\begin{eqnarray}
 f_{\rm b} &=& \sum_{[m,n]\leq B(k)} f_{m,n}
 \big( \delta_{m,k-n}-3\delta_{m,k-n-1}+3\delta_{m,k-n-2}-\delta_{m,k-n-3} \big)
 \nonumber \,, \\
 f_{\rm s} &=& \sum_{[m,n]\leq B(k)} f_{m,n}
 \big((1-m)\delta_{m,k-n}+(3m-1)\delta_{m,k-n-1}
 \,, \nonumber \\
 & & \qquad -(3m+1)\delta_{m,k-n-2} + (m+1)\delta_{m,k-n-3} \big)
 \label{Entingformulae} \\
 f_{\rm c} &=& \sum_{[m,n]\leq B(k)} f_{m,n}
 \big((m-1)(n-1)\delta_{m,k-n}+(1+m+n-3mn)\delta_{m,k-n-1}
 \nonumber \\
 & & \qquad +(3mn+m+n-1)\delta_{m,k-n-2}-(m+1)(n+1)\delta_{m,k-n-3} \big)
 \nonumber \,,
\end{eqnarray}
where all the $f_{m,n}$ with $m$ or $n$ non-positive are zero by
convention. Compared to Enting's formulae, the expression for $f_{\rm s}$ has
been simplified under the assumption that the $f_{m,n}$ are
symmetric in $m$ and $n$, which will always be the case in this
paper.

\subsubsection{Triangular lattices in triangles.}
\label{sec:FLM_triangle}

For the sake of studying models defined on the triangular lattice, it
is sometimes useful to work with only one type of corners. One then
seeks the free energy $f_M$ of the model defined on a large equilateral
triangle of side $M$, as shown in Fig.~\ref{lattices}c.
We have derived the corresponding FLM formulae,
going through the same steps as above.

The starting point is now
\begin{equation}
 f_{M}\approx \sum_{i\leq k}{\frac{(M-i+1)(M-i+1)}{2}\tilde{f}_{i}} \,,
 \label{fMapprox}
\end{equation}
where the self-consistency condition
\begin{equation}
 f_{m}= \sum_{i\leq m}{\frac{(m-i+1)(m-i+1)}{2}\tilde{f}_{i}} 
 \label{fmselfcons}
\end{equation}
yields
\begin{equation}
 \tilde{f}_{i}= \sum_{m\leq i}f_m (\delta_{m,i}-3\delta_{m,i-1}
 +3\delta_{m,i-2}-\delta_{m,i-3}) \,.
\end{equation}

Finally, the free energy is written as 
\begin{equation}
 f_M = \frac{M(M+1)}{2} f_{\rm b} + M f_{\rm s} + f_{\rm c} \,,
\label{fM_decomposition}
\end{equation}
where the different contributions are given by
\begin{eqnarray}
 f_{\rm b} &=& \sum_{m\leq k} f_{m}
 \big( \delta_{m,k}-2\delta_{m,k-1}+\delta_{m,k-2} \big)
 \nonumber \,, \\
 f_{\rm s} &=& \sum_{m\leq k} f_{m}
 \big( (1-m)\delta_{m,k}+(2m+1)\delta_{m,k-1}-(m+2)\delta_{m,k-2} \big)
 \,, \label{Entingformulaetriang} \\
 f_{\rm c} &=& \frac{1}{2}\sum_{m\leq k} f_{m}
 \big( (m^2-3m+2)\delta_{m,k}+2(1-m^2)\delta_{m,k-1}
 \nonumber \\
 & & \qquad
 +(m^2+3m+2)\delta_{m,k-2} \big)
 \,. \nonumber
\end{eqnarray}

\subsection{Results}

\subsubsection{Selfdual Potts model on the square lattice.}
\label{sec:selfdual_potts_square}

On its selfdual curve, the square-lattice Potts model is equivalent to
a loop model, where each loop carries a statistical weight
$n=\sqrt{Q}$. The partition functions of finite lattices of size
$M\times N$ reads
\begin{equation}
 Z_{M,N}=n^{M N} P_d(n^{-1}) \,,
\end{equation}
where $P_d(x)$ is a polynomial of degree $d=M N - 1$. The constant
coefficient $P_d(0)=1$ corresponds to the ground state shown in
Fig.~\ref{ground_state}a.

The FLM formulae thus reduce to an expansion in powers of $1/n$, which
is correct up to an order determined by the size of the graph with
highest weight that cannot fit into the cutoff set $B(k)$. Compared to
the ground state configuration, this property is satisfied by the
graphs in which the edge subset $A \subset E$ of (\ref{Z_Potts_FK})
forms one straight or L-shaped figure of total length $k+1$, which
contribute to the order $n^{-k}$. We thus expect the FLM formulae
(\ref{Entingformulae}) with a cutoff $B(k)$ to give
an approximation of the $M \times N$ free
energy that is correct up to order $n^{-k}$.

Using a transfer matrix algorithm, we have computed the polynomials
$P_d(x)$ for all finite lattices within the cutoff set $B(31)$,
allowing correct results up to order $n^{-31}$.

Using instead of $1/n$ the small parameter $q$ defined by 
\begin{equation}
 n=q+\frac{1}{q}
\end{equation}
the exponentiated free energies $\rme^{f_{\rm b}}$, $\rme^{f_{\rm s}}$
and $\rme^{f_{\rm c}}$ finally turn out to have up to this order the
following nice expressions
\begin{eqnarray}
  \rme^{f_{\rm b}} &=& \frac{q^2+1}{q^2(q-1)^2}\prod_{k=1}^{\infty}{\left(\frac{1-q^{4k-1}}{1-q^{4k+1}}\right)^4} \,, \nonumber \\
  \rme^{f_{\rm s}} &=& (1-q)\prod_{k=1}^{\infty}{\left(\frac{1-q^{8k-1}}{1-q^{8k-5}}\right)^2} \,, \label{products selfdualsquare} \\
  \rme^{f_{\rm c}} &=&  \prod_{k=1}^{\infty}{\frac{1}{(1-q^{8k-6})(1-q^{8k-4})^4(1-q^{8k-2})}} \,. \nonumber
\end{eqnarray}

The expression for $\rme^{f_{\rm b}}$ is shown in \ref{appendix comparison
  products analytical} to be equivalent to the analytical expression
given by Baxter in \cite{BaxterBook}, that is
\begin{equation}
 \psi = -\frac{1}{2} \ln Q - 2\left[\beta + \sum_{k=1}^{\infty}{\frac{1}{k}\rme^{-k\lambda} \frac{\sinh 2 k \beta}{\cosh  k \lambda} } \right] \,,
\label{BaxterPottsSquareBulk}
\end{equation}
where $\psi \equiv -\lim_{N\to\infty} \frac{1}{N} \ln Z_N$ is defined as the
dimensionless free energy in the thermodynamic limit (here $Z_N$ denotes
the partition function of a system with $N$ spins).
The correspondence with our notation is
\begin{eqnarray}
 \psi &=& -f_{\rm b} \,, \nonumber \\
 Q^{1/2} &=& 2\cosh\lambda \,, \nonumber \\
 \frac{v}{Q^{1/2}} &=& \frac{\sinh\beta}{\sinh(\beta-\lambda)} \,. \nonumber
\end{eqnarray}

The regular form exhibited by the product expression for $\rme^{f_{\rm
    b}}$---which is proved starting from Baxter's result
(\ref{BaxterPottsSquareBulk}) in \ref{appendix_A1}---and its
structural similarity with the corresponding conjectured expressions
for $\rme^{f_{\rm s}}$ and $\rme^{f_{\rm c}}$ encourages us to assume
that the two latter expressions are also valid to all orders.  These
expressions are, as far as we know, new.%
\footnote{But in the case of $\rme^{f_{\rm s}}$ it seems likely that
  an equivalent result exists somewhere in the literature.}

In general we expect $\rme^{f_{\rm b}}$, $\rme^{f_{\rm s}}$ and
$\rme^{f_{\rm c}}$ for {\em all} the models studied in this paper to
have an expression of the form (\ref{generic factorized form}), modulo
some simple overall factors such as $\frac{q^2+1}{q^2(q-1)^2}$ or
$(1-q)$ appearing in (\ref{products selfdualsquare}).  Moreover, we
expect the coefficients $\beta_k$ and $\gamma_k$ to be {\em periodic}
functions of $k$.  For instance the results found in (\ref{products
  selfdualsquare}) correspond to the simplest case of $\beta_k = 0$;
and the periodicities are $4$ for $\rme^{f_{\rm b}}$, and $8$ for
$\rme^{f_{\rm s}}$ and $\rme^{f_{\rm c}}$. We shall discuss the issue
of periodicity further in section~\ref{sectionperiodicities}.

The above assumption will be further corroborated by the study of
other models in the remainder of the paper.

We remark that a result such as $\rme^{f_{\rm c}}$ in (\ref{products
  selfdualsquare}) can be said to be established beyond any reasonable
doubt, since the coefficients $\gamma_0,\gamma_1,\ldots,\gamma_{31}$
determined numerically cover four complete periods. Once the
periodicity property is established (or assumed) the determination of
the coefficients over a single period (or a little more to be sure)
would be sufficient.

One more important feature of the generic form (\ref{generic
  factorized form}) is that it gives a particular meaning to the limit
$q\to 1^{-}$, which turns out to be critical for all the considered
models. The convergence of products of the type (\ref{generic
  factorized form}) when $q\to 1^{-}$ indeed depends in a non-obvious
way on the coefficients $\alpha_k$, giving rise to different critical
behaviours to be studied in the following.

\subsubsection{Antiferromagnetic Potts model on the square lattice.}

The antiferromagnetic transition curve \cite{AFBaxter} for the square-lattice
Potts model is given by (\ref{AFPottscritical}). To resolve the square root
we parameterise $Q=-(q^{-1}-q)^2$, with $q$ taking real (resp.\ complex) values
in the non-critical (resp.\ critical) regime. We have then $v = (q^{1/2}-q^{-1/2})^2$.
This allows us to compute the partition function of finite lattices as polynomials
in $q$. More precisely, the partition functions for lattices of size $M\times N$ read
\begin{equation}
 Z_{MN} = \frac{q^{M+N-3MN}}{(q^{-1}-q)^{MN}} P_d(q) \,,
\end{equation}
where $P_d(q)$ is a polynomial of degree $d=3MN-M-N$ in $q$ with constant
coefficient $P_d(0)=1$. To obtain $P_d$, FLM calculations are applied in exactly
the same way as for the selfdual curve. The numerical calculations are now done
up to order $q^{22}$, i.e., using the finite partition functions with the cutoff set
$B(23)$.
This gives for the bulk, surface and corner and free energies:
\begin{eqnarray}
 \rme^{f_{\rm b}} &=& -\frac{(1-q)^2(1-q^2)}{q^2}
 \prod_{k=1}^{\infty}{\left(\frac{1-q^{8k-2}}{1-q^{8k-6}}\right)^2} 
 \,, \nonumber \\
 \rme^{f_{\rm s}} &=& \frac{1}{(1-q)}
 \prod_{k=1}^{\infty}{\frac{(1-q^{16k-14})(1-q^{16k-2})}{(1-q^{16k-10})(1-q^{16k-6})}} 
 \prod_{k=1}^{\infty}{\left(\frac{1-q^{8k-3}}{1-q^{8k+1}}\right)^2}
 \,, \\
 \rme^{f_{\rm c}} &=&
 \prod_{k=1}^{\infty}\frac{(1-q^{4k-2})(1-q^{8k-4})^3}{(1-q^{16k-8})^4} 
 \,. \nonumber 
\label{productsAF}
\end{eqnarray}

The presence of an overall minus sign in ${\rm e}^{f_{\rm b}}$ stems
from the fact that $Q<0$, and should not be taken into account as for
the critical properties of the model when $Q\to 0$. The expression
found for $\rme^{f_{\rm b}}$ is shown in \ref{appendix_A2} to be
equivalent to an expression given by Baxter in section 12.5 of \cite{AFBaxter}, up to
an analytic continuation. More precisely, using the same notations as
for the self-dual case, the free energy $\psi$ is given for $Q\geq 4$
by (\ref{BaxterPottsSquareBulk}), where this time $\beta$ has
to be replaced by a parameter $u$ defined by
\begin{equation}
 {\rm e}^K = \frac{\sinh(\lambda + u)}{\sinh(\lambda - u)} \,,
 \qquad 0\leq \Im u < \pi \,,
\end{equation}
the expression being valid under the condition $0 < \Re u < \lambda$.
On the contrary, the expressions for $\rme^{f_{\rm s}}$ and
$\rme^{f_{\rm c}}$ are new.

\subsubsection{Selfdual Potts model on the triangular lattice.}
\label{triangular Potts}

The selfdual transition curve on the triangular lattice is given by the cubic
(\ref{triangularPottscritical}). It can be resolved through the parameterisation
\begin{eqnarray}
 \sqrt{Q} &=& t^{3/2} + t^{-3/2}
 \,, \nonumber \\
 v &=& -1 + t + t^{-1} \,.
\end{eqnarray}
The FLM calculations are then conducted in terms of the variable $t=q^{2/3}$.

The partition functions read
\begin{equation}
 Z_{MN} = \frac{(1+t)^2(1-t+t^2)^{M N + 1}}{t^{3MN}} P_d(t) \,,
\end{equation}
where $P_d(t)$ is a polynomial of degree $d=8(MN-1)$ in $t$ with
constant coefficient $P_d(0)=1$. Finite lattice calculations similar
to those used so far are then applied to obtain the $P_d$. The only
difference is that this time the highest order graph that cannot fit
into the cutoff set $B(k)$ is the straight line graph oriented
diagonally, formed by $\sim k/2$ diagonal edges, which contributes to
order $t^{k}$ in the partition function. We write the final results in
terms of the same variable $q$ as used in the square-lattice case.

The numerical calculations are conducted up to order $t^{23}$ (i.e.,
with cutoff set $B(23)$), that is to order $q^{46/3}$. They lead to
the following product expressions for the bulk and surface free
energies:
\begin{eqnarray}
 \rme^{f_{\rm b}} &=& \frac{1}{q^2}\frac{1-q^4}{1-q^2}
 \prod_{k=1}^{\infty} {\left(\frac{(1-q^{4k-\frac{4}{3}})(1-q^{4k-\frac{2}{3}})}
 {(1-q^{4k-\frac{8}{3}})(1-q^{4k+\frac{2}{3}})}\right)^3}
  \,, \nonumber \\
 \rme^{f_{\rm s}} &=& \frac{(1-q^{\frac{4}{3}})^2}{(1-q^{\frac{2}{3}})^2} 
 \prod_{k=1}^{\infty} {\left(\frac{(1-q^{8k-\frac{8}{3}})(1-q^{8k-\frac{22}{3}})}
 {(1-q^{8k-\frac{11}{3}})(1-q^{8k-\frac{14}{3}})}\right)^2}
 \,. \label{efs_tri_sd}
\end{eqnarray}

The expression (\ref{efs_tri_sd}) of $\rme^{f_{\rm b}}$ is shown in
\ref{appendix_A3} to be equivalent to an expression given by Baxter in section 12.6 of 
\cite{BaxterBook}, that is,
\begin{equation}
 \psi = -\frac{1}{2} \ln Q - 3\left[\beta + \sum_{k=1}^{\infty}{\frac{1}{k}\rme^{-k\lambda} \frac{\sinh 2 k \beta}{\cosh  k \lambda} } \right] \,,
\label{BaxterPottsTri2}
\end{equation}
with the same notations as in section
\ref{sec:selfdual_potts_square}. The expression found for
$\rme^{f_{\rm s}}$ seems, on the contrary, new.

The series expansion of $\rme^{f_{\rm c}}$ appears to be more complicated.
Up to order $q^{46/3}$ we find: 
\begin{eqnarray}
 \rme^{f_{\rm c}} &=&  (1-q^{\frac{4}{3}})^{-1}(1-q^2)^{-1}(1-q^{\frac{8}{3}})
 (1-q^{\frac{10}{3}})^{-1}(1-q^4)^{-2}(1-q^{\frac{14}{3}})^{-1}
 \nonumber \\
 & & (1-q^{\frac{16}{3}})^{-5}(1-q^{6})^{5}(1-q^{\frac{20}{3}})^{-9}(1-q^{\frac{22}{3}})^{9}
 (1-q^{8})^{-8}(1-q^{\frac{26}{3}})^{9}
 \nonumber \\
 & & (1-q^{\frac{28}{3}})^{-15}(1-q^{10})^{15}(1-q^{\frac{32}{3}})^{-11}
 (1-q^{\frac{34}{3}})^{13}(1-q^{12})^{-22}
 \nonumber \\
 & & (1-q^{\frac{38}{3}})^{19} (1-q^{\frac{40}{3}})^{-17}(1-q^{14})^{19}(1-q^{\frac{44}{3}})^{-29}  + \mathcal{O}(q^{\frac{46}{3}}) \,.
 \label{efc_tri_sd}
\end{eqnarray}
We believe that the correct expression is still of the form
(\ref{generic factorized form}), but for the first time $\beta_k \neq
0$.

We recall that (\ref{efc_tri_sd}) applies to the triangular lattice
inscribed in a rectangle, cf.\ Fig.~\ref{lattices}b. This has two
different types of corners: 1) two of angle $\frac{2\pi}{3}$ each
connected to three edges, and 2) two of angle $\frac{\pi}{3}$ each
connected to two edges. The contribution from each type of corner to
$\rme^{f_{\rm c}}$ is expected to be different. Therefore we should be
looking for an expansion of the form
\begin{equation}
 \rme^{f_{\rm c}} = \prod_{k=1}^\infty {(1-q^k)^{\alpha_k^{(1)}+\alpha_k^{(2)}}} \,,
\end{equation}
where $\alpha_k^{(i)} = \beta_k^{(i)} k+\gamma_k^{(i)}$. In accordance
with what was observed for previous models, we expect the exponents
$\beta_k^{(i)}$ and $\gamma_k^{(i)}$ to exhibit the same periodicity
as in the expression for $\rme^{f_{\rm s}}$; see
(\ref{efs_tri_sd}). This periodicity is $8$ in the $q$ variable, but
since powers of $q^{2/3}$ occur we need 12 consecutive terms to
determine one period. Moreover there are now $4$ different types of
exponents, so the complete determination would require a cutoff set at
least $B(48)$, which is clearly out of reach of the methods used this
far.

To proceed we may turn to lattices inscribed in a triangle [cf.\
Fig.~\ref{lattices}c], for which we expect the bulk and surface free
energy to be unchanged, but which possess three corners of the {\em
  same} type, namely of angle $\frac{\pi}{3}$ (type $i=2$).  Computing
finite-lattice partition functions by a transfer matrix algorithm
similar to those used previously, and using the FLM formulae of
section~\ref{sec:FLM_triangle}, yields indeed the same expressions as
above for $\rme^{f_{\rm b}}$ and $\rme^{f_{\rm s}}$ up to order
$t^{18}$, except from the fact that the exponents $\alpha_k=2$ in
(\ref{efs_tri_sd}) are now replaced with $\alpha_k=3$, which stems
from the different definitions (\ref{fMN_decomposition}) and
(\ref{fM_decomposition}) chosen for $f_{\rm s}$ in the rectangle and
in the triangle case. For the corner free energy we find up to this
order
\begin{equation}
 \rme^{f_{\rm c}} = \prod_{k=1}^{\infty}\frac{1}{(1-q^{8k-6})(1-q^{8k-2})
 (1-q^{8k-14/3})^3(1-q^{8k-10/3})^3} \,.
\label{efc_Potts_tri_tri} 
\end{equation}
Quite remarkably, this expression shows that $\beta_k^{(2)} = 0$, so
the linear growth of the exponents visible in (\ref{efc_tri_sd}) must
be due to $\beta_k^{(1)} \neq 0$. Insight can thus be gained by
factorising out the contribution of two corners of angle
$\frac{\pi}{3}$ [that is, eq.~(\ref{efc_Potts_tri_tri}) to the
power $\frac{2}{3}$] from (\ref{efc_tri_sd}). The remainder corresponds
to the contribution of two corners of angle $\frac{2\pi}{3}$, and is
given by
\begin{eqnarray}
 & \left[\prod_{k=1}^{\infty}{(1-q^{8k-6}) (1-q^{8k-2})}\right]^{2/3} \times
   (1-q^{\frac{4}{3}})^{-1} (1-q^2)^{-1} (1-q^{\frac{8}{3}}) \nonumber \\
 & (1-q^{\frac{10}{3}})^{-1} (1-q^4)^{-2} (1-q^{\frac{14}{3}})^{3}
   (1-q^{\frac{16}{3}})^{-5} (1-q^{6})^{5} (1-q^{\frac{20}{3}})^{-9} \nonumber \\
 & (1-q^{\frac{22}{3}})^{9} (1-q^{8})^{-8} (1-q^{\frac{26}{3}})^{9}
   (1-q^{\frac{28}{3}})^{-15} (1-q^{10})^{15} (1-q^{\frac{32}{3}})^{-11}
   \nonumber \\
 & (1-q^{\frac{34}{3}})^{15} (1-q^{12})^{-22} (1-q^{\frac{38}{3}})^{21}
   (1-q^{\frac{40}{3}})^{-17} (1-q^{14})^{19} (1-q^{\frac{44}{3}})^{-29}
   \nonumber \\
 & +\mathcal{O}(q^{\frac{46}{3}}) \,.
\end{eqnarray}
Unfortunately this is not quite enough in order to be able to
conjecture the corresponding exact expression. According to the above
discussion, we would need the first 24 exponents (and ideally a few
more as a verification), but we only have 22. For the sake of checking
consistency, we tried to adapt the FLM formulas to lattices inscribed
in hexagons, that is, lattices with six corners of type
$i=1$. However, it seems that in this case the FLM formulae are
impossible to invert.

\subsubsection{Chromatic polynomial on the triangular lattice.}
\label{sec:chrom_poly_tri}

The finite partition functions for the $Q$-colour chromatic polynomial
on a triangular lattice inscribed in a rectangle were calculated by
one of the authors in \cite{JacobsenChromatic}. This resulted in the
following product formulae for the bulk, surface and corner free
energies:
\begin{eqnarray}
 \rme^{f_{\rm b}} &=& -\frac{1}{x}\prod_{k=1}^{\infty}{\frac{(1-x^{6k-3})(1-x^{6k-2})^2(1-x^{6k-1})}{(1-x^{6k-5})(1-x^{6k-4})(1-x^{6k})(1-x^{6k+1})}}
 \,, \nonumber \\
 \rme^{f_{\rm s}} &=& \prod_{k=1}^{\infty}{\left(\frac{(1-x^{12k-11})(1-x^{12k-6})(1-x^{12k-5})(1-x^{12k-4})}{(1-x^{12k-9})(1-x^{12k-8})(1-x^{12k-7})(1-x^{12k-2})}\right)^2}
 \,,  \nonumber \\
 \rme^{f_{\rm c}} &=& \prod_{k=1}^{\infty}{(1-x^{12k-11})^{-1}(1-x^{12k-10})(1-x^{12k-9})^{3}(1-x^{12k-8})^{2}}  \nonumber \\ 
  & & \qquad (1-x^{12k-7})^{4}(1-x^{12k-6})^{-3}(1-x^{12k-5})(1-x^{12k-4})^{-3}  \nonumber \\ 
 & & \qquad (1-x^{12k-3})^{5}(1-x^{12k-2})^{2}(1-x^{12k-1})^{2}(1-x^{12k})  \,, \label{efc chromatic rectangle}
\end{eqnarray}
where the expansion parameter $x$ is defined this time by
$Q=2-x-\frac{1}{x}$.

As for the selfdual Potts model on the triangular lattice, these
expressions can be checked by FLM calculations on the triangular
lattice inscribed in a triangle.  We have computed the corresponding
finite-lattice partition functions up to triangles of size $M=21$,
which gives access to the bulk, surface and corner free energy series
up to order $\sim x^{20}$. The expressions for $\rme^{f_{\rm b}}$ and
$\rme^{f_{\rm s}}$ coincide with (\ref{efc chromatic rectangle}) as
expected, whereas we find for the corner free energy
\begin{eqnarray}
 \rme^{f_{\rm c}} &=& \prod_{k=1}^{\infty} (1-x^{12k-10}) (1-x^{12k-9})^2 (1-x^{12k-8})
 (1-x^{12k-7})^3
 \nonumber \\
 & & \qquad (1-x^{12k-6})^{-2} (1-x^{12k-5})^3 (1-x^{12k-4})^{-2}(1-x^{12k-3})^3
 \nonumber \\
 & & \qquad (1-x^{12k-2})(1-x^{12k-1})^3(1-x^{12k}) \,.
 \label{efc chromatic triangle}
\end{eqnarray}

We can now combine eq.~(\ref{efc chromatic rectangle}) for two corners
of type $i=1$ and two corners of type $i=2$ with eq.~(\ref{efc
  chromatic triangle}) for three corners of type $i=2$, to obtain the
exact formulae for a single corner of each type:
\begin{eqnarray}
 \rme^{f_{\rm c}(\frac{2\pi}{3})} &=& \prod_{k=1}^{\infty}
 (1-x^{12k-11})^{-1/2} (1-x^{12k-10})^{1/6} (1-x^{12k-9})^{5/6}
 \nonumber \\
 & & (1-x^{12k-8})^{2/3} (1-x^{12k-7}) (1-x^{12k-6})^{-5/6}
 \nonumber \\
 & & (1-x^{12k-5})^{-1/2} (1-x^{12k-4})^{-5/6} (1-x^{12k-3})^{3/2}
 \nonumber \\
 & & (1-x^{12k-2})^{2/3} (1-x^{12k})^{1/6} \,, \\
 \rme^{f_{\rm c}(\frac{\pi}{3})} &=& \prod_{k=1}^{\infty}
 (1-x^{12k-10})^{1/3} (1-x^{12k-9})^{2/3} (1-x^{12k-8})^{1/3}
 \nonumber \\
 & & (1-x^{12k-7}) (1-x^{12k-6})^{-2/3} (1-x^{12k-5})
 \nonumber \\
 & & (1-x^{12k-4})^{-2/3} (1-x^{12k-3}) (1-x^{12k-2})^{1/3}
 \nonumber \\
 & & (1-x^{12k-1}) (1-x^{12k})^{1/3} \,.
\end{eqnarray}

The chromatic polynomial on the triangular lattice is critical in the
region $Q\in[0,4]$.  As for the selfdual Potts model, we are
interested in the limit $Q\to 4^{+}$, which we parameterise in terms
of the usual variable $q$. It is related to the above variable $x$ by
$x=-q^2$. To turn the above products into expressions in terms of $q$
thus amounts simply to the following replacements,
\begin{eqnarray}
1-x^{2p} & \qquad \mbox{is replaced by}  \qquad 1-q^{4p} \,, \nonumber \\
1-x^{2p+1} & \qquad \mbox{is replaced by} \qquad \frac{1-q^{8p+4}}{1-q^{4p+2}} \,.
 \label{x to q}
\end{eqnarray}
We will also be interested in the limit $Q\to 0^-$, which can be
recovered as $q\to \mathrm{i} \times 1^-$, i.e., using complex values
of $q$.

\subsubsection{${\rm FPL}^2$ model.}

For lattices of the type shown in Fig.~\ref{fig:FPL2_lattice} the
partition function is
\begin{equation}
 Z_{M N} = 2 n^{2 M N + M + N} P_d(n^{-2}) \,,
 \label{Z_FPL2_finite}
\end{equation}
where $P_d(x)$ is a polynomial of degree $d = M N + \lfloor (M+N)/2
\rfloor - 1$ in the variable $x = n^{-2}$ with constant coefficient
$P_d(0) = 1$. The overall factor of $2$ in (\ref{Z_FPL2_finite}) is
unimportant for what follows, and it is convenient to regard the power
of $n$ as a multiplicative contribution of $n^2$ to ${\rm e}^{f_{\rm
    b}}$ and a contribution of $n$ to ${\rm e}^{f_{\rm s}}$.

We have computed the polynomials $P_d$ in (\ref{Z_FPL2_finite}) for
all lattices with $M+N \le 17$. Introducing the usual parameterisation
$n = q + q^{-1}$, and using the FLM, this yields series expansions in
powers of $q$ for the the bulk, surface and corner free energies,
${\rm e}^{f_{\rm b}}$, ${\rm e}^{f_{\rm s}}$ and ${\rm e}^{f_{\rm c}}$,
that are correct up to order $q^{22}$. Since $P_d$ depends on
$n^{-2}$, only even powers of $q$ appear in these series expansions.

Exact product formulae for these series can readily be conjectured.
Reinstating the above-mentioned multiplicative contributions we obtain
\begin{eqnarray}
\rme^{f_{\rm b}} &=& \frac{1}{q^2}\prod_{k=1}^{\infty}{\left(\frac{(1-q^{8k-4})(1-q^{8k-2})}{(1-q^{8k-6})(1-q^{8k})}\right)^4}  \,, \nonumber \\
\rme^{f_{\rm s}} &=& \frac{(1-q^2)^2}{q} \prod_{k=1}^{\infty}{\left(\frac{(1-q^{16k-12})(1-q^{16k-10})(1-q^{16k-2})^3}{(1-q^{16k-14})^2(1-q^{16k-6})^2(1-q^{16k-4})}\right)^2}   \,, \\
\rme^{f_{\rm c}} &=& \prod_{k=1}^{\infty} \frac{1}{(1-q^{16k-14})^5(1-q^{16k-12})(1-q^{16k-10})} \times \nonumber \\
 & & \qquad \frac{1}{(1-q^{16k-6})^5(1-q^{16k-4})^5(1-q^{16k-2})} \,. \nonumber 
\end{eqnarray}

Note that the above expression for $\rme^{f_{\rm b}}$ is identical to
the one obtained in Eq.~(64) of \cite{DeiCont1} from an exact Bethe
Ansatz solution.  On the contrary, the expressions found for
$\rme^{f_{\rm s}}$ and $\rme^{f_{\rm c}}$ are new.

\subsubsection{Ising model on the square lattice.}

Turning to the case of the two-dimensional Ising model on the square
lattice with nearest-neighbour coupling $J$, the partition function is
calculated from the FLM formulae as a low-temperature expansion in
powers of the variable $x = {\rm e}^{-K} = {\rm e}^{-\beta J}$. The
partition functions read
\begin{equation}
Z_{MN} = \frac{1}{x^{MN-M-N}}P_d(x) \,,
\end{equation}
where $P_d(x)$ is a polynomial of degree $d=2(2MN-M-N)$ in $x$ with
constant coefficient $P_d(0)=1$. From the same arguments as for the
selfdual Potts model, the use of a cutoff set $B(k)$ leads to an
approximation that is correct to order $\sim x^{2k}$. The choice of a ``good''
variable in terms of which the bulk, surface and corner free energies
take a well-behaved factorised form is inspired by Baxter, Sykes and
Watts \cite{BaxterIsing}. We therefore set
\begin{equation}
 x^2 = q^{1/2}\prod_{k=1}^{\infty}
 {\frac{(1-q^{8k-7})(1-q^{8k-1})}{(1-q^{8k-5})(1-q^{8k-3})}} \,.
\label{eq: Ising}
\end{equation}
One can verify that the critical point $\beta_{\rm c}
J=\frac{1}{2}\log(1+\sqrt{2})$ corresponds to the limit $q\to1^-$.  In
terms of the variable $q$ we find, using the cutoff set $B(49)$ (i.e.,
up to order $q^{22}$), that
\begin{eqnarray}
 {\rm e}^{f_{\rm b}} &=& \frac{1}{q^{1/2}} \prod_{k=1}^{\infty}
 \frac{(1-q^{8k-1})^{8k-1}(1-q^{8k-5})^{8k-5}}
      {(1-q^{8k-7})^{8k-7}(1-q^{8k-3})^{8k-3}}
 \frac{(1-q^{8k-4})^{2}}
      {(1-q^{8k-6})(1-q^{8k-2})} \,, \nonumber \\
 {\rm e}^{f_{\rm s}} &=& x^{-1} \prod_{k=1}^{\infty} \left(
 \frac{1-q^{\frac{8k-3}{2}}}
      {1-q^{\frac{8k-5}{2}}} \right)^{4k-2} \left(
 \frac{1-q^{\frac{8k-1}{2}}}
      {1-q^{\frac{8k+1}{2}}} \right)^{4k} \nonumber \\ 
 & &  \qquad \quad \left(
 \frac{1-q^{8k-5}}
      {1-q^{8k-3}} \right)^{2k-1} \left(
 \frac{1-q^{8k+1}}
      {1-q^{8k-1}} \right)^{2k} \,, \label{efb_Isingsquare} \\
 {\rm e}^{f_{\rm c}} &=& \prod_{k=1}^{\infty}
 \frac{1}{1-q^{8k-4}}
 \frac{(1-q^{8k-6})^{4k}}
      {(1-q^{8k-2})^{4k-4}} \left(
 \frac{1-q^{4k-1}}{1-q^{4k-3}} \right)^{8k-4} \nonumber \,. 
\end{eqnarray}

The expression found for $\rme^{f_{\rm b}}$ is shown in
\ref{appendix_A4} to be equivalent to the analytical expression given
in section 11.8 of \cite{BaxterBook} for the dimensionless free energy in the
thermodynamical limit, namely
\begin{equation}
\psi = -2K-\tau+g(z)+g(z^{-1}) \,,
\end{equation}
where $K = \beta J$ is the dimensionless Ising coupling, $z=1$ at the
critical point,
\begin{equation}
g(z) = \sum_{m=1}^{\infty}{\frac{q^{3m}(1-q^{2m})(z^{-m}-q^{2m}z^m)}{m(1-q^{8m})(1+q^{2m})}} \,,
\end{equation}
and $\tau$ can be expressed as a function of $q$ as
\begin{equation}
\tau = \sum_{m=1}^{\infty}{\frac{q^{2m}(1-q^{2m})^2}{m(1-q^{8m})}} \,.
\end{equation}
Another equivalent expression was given in \cite{BaxterIsing}, namely
\begin{equation}
-\psi = 2K + \sum_{n=1}^{\infty}{\frac{q^{2n}(1-q^{n})^2(1-q^{2n})^2}{n(1+q^{2n})(1-q^{8n})}} \,.
 \label{fb Ising analytical}
\end{equation}

\subsubsection{Ising model on the triangular lattice.}

In the case of the Ising model on a triangular lattice, the natural
variable to be used for product expansions is inspired by section 11.8
of Baxter's book \cite{BaxterBook}, namely
\begin{equation}
 x^2 = q^{\frac{1}{3}} \prod_{k=1}^{\infty}
 \frac{(1-q^{8k-7+\frac{1}{3}})(1-q^{8k-1-\frac{1}{3}})}
 {(1-q^{8k-5-\frac{1}{3}})(1-q^{8k-3+\frac{1}{3}})} \,.
 \label{param_x_Ising_tri}
\end{equation}
The critical point $K_c \equiv \beta_c J=\frac{1}{4}\log 3$
corresponds once again to the limit $q\to 1^-$.

The finite-lattice partition functions read 
\begin{equation}
  Z_{MN} = \frac{2}{x^{3MN-2M-3N+1}} P_d(x^2) \,,
\end{equation}
where $P_d(x^2)$ is a polynomial of degree $d=2MN-M-N$ in $x^2$ with
constant coefficient $P_d(0)=1$. From the same arguments as for the
triangular-lattice Potts model, the highest order graph that cannot
fit into the FLM cutoff set $B(k)$ is a diagonally oriented line graph
of length $\sim k/2$, corresponding to inverting $\sim k$ surrounding
links and thus contributing to the partition function up to an order
$\sim x^{2k}$. Leading calculations up to the cutoff set $B(53)$, we
thus find up to order $q^{50/3}$ for the bulk free energy
\begin{eqnarray}
 {\rm e}^{f_{\rm b}} &=&  x^{-3} \prod_{k=1}^{\infty}
 \frac{(1-q^{8k-4})^{2}}
      {(1-q^{8k-6})(1-q^{8k-2})} \prod_{k=1}^{\infty} \left(
 \frac{1-q^{8k-\frac{14}{3}}}
      {1-q^{8k-\frac{10}{3}}} \right)^{6k-1} \nonumber \\    
 & & \qquad \prod_{k=1}^{\infty} \left( 
 \frac{(1-q^{8k-\frac{4}{3}})(1-q^{8k-\frac{2}{3}})(1-q^{8k+\frac{8}{3}})}
      {(1-q^{8k-\frac{8}{3}})(1-q^{8k+\frac{2}{3}})(1-q^{8k+\frac{4}{3}})}
 \right)^{6k} \,.
\label{product_efb_ising_tri}
\end{eqnarray}
This expression is shown in \ref{appendix_A5} to be equivalent to the
analytical expression given in \cite{BaxterBook} for the dimensionless
free energy in the thermodynamic limit, namely
\begin{equation}
\psi = -3K - \tau + 3g(z) \,,
\label{analytical_efb_ising_tri}
\end{equation}
with the same notations as on the square lattice, and this time
$z=q^{-1/3}$ at the critical point.  Another equivalent analytical
expression was given by Baxter, Sykes and Watts in \cite{BaxterIsing}.

We were unable to obtain a regular factorised form for the surface and
corner energies, for the same reasons as those given in
section~\ref{triangular Potts} in the case of the triangular-lattice
Potts model. We nevertheless give hereafter the first terms of the
factorised series expansion in terms of the variable $q$:
\begin{eqnarray}
 {\rm e}^{f_{\rm s}} &=&  x^2
 (1-q^\frac{4}{3})^{-2} (1-q^\frac{8}{3})^4 (1-q^\frac{10}{3})^2
 (1-q^\frac{14}{3})^{-4} (1-q^\frac{16}{3})^{-4} \nonumber \\
 & & (1-q^\frac{20}{3})^2 (1-q^\frac{22}{3})^6 (1-q^\frac{26}{3})^{-6}
 (1-q^\frac{28}{3})^{-6} (1-q^\frac{32}{3})^{12} \nonumber \\
 & & (1-q^\frac{34}{3})^8 (1-q^\frac{38}{3})^{-10} (1-q^\frac{40}{3})^{-12}
 (1-q^\frac{44}{3})^6 (1-q^\frac{46}{3})^{12}
 + \mathcal{O}(q^{\frac{50}{3}}) \nonumber \\
 {\rm e}^{f_{\rm c}} &=& \frac{2}{x}
 (1-q^\frac{2}{3})^{-2} (1-q)^{-2} (1-q^\frac{4}{3})^{4}
 (1-q^\frac{5}{3})^{-4} (1-q^2)^{6} \nonumber \\
 & & (1-q^\frac{7}{3})^{-4} (1-q^\frac{8}{3})^{4} (1-q^3)^{-2}
 (1-q^\frac{10}{3})^{3} (1-q^4)^{-3} \nonumber \\
 & & (1-q^\frac{14}{3})^{-1} (1-q^5)^{-2} (1-q^\frac{16}{3})^{-4}
 (1-q^\frac{17}{3})^{-4} (1-q^6)^{4} \nonumber \\
 & & (1-q^\frac{19}{3})^{-4} (1-q^\frac{20}{3})^{10} (1-q^7)^{-2}
 (1-q^\frac{22}{3})^{6} (1-q^8)^{-2} \nonumber \\
 & & (1-q^\frac{26}{3})^{-10} (1-q^9)^{-2} (1-q^\frac{28}{3})^{2}
 (1-q^\frac{29}{3})^{-4} (1-q^{10})^{8} \nonumber \\
 & & (1-q^\frac{31}{3})^{-4} (1-q^\frac{32}{3})^{-8} (1-q^{11})^{-2}
 (1-q^\frac{34}{3})^{7} (1-q^{12}) (1-q^\frac{38}{3})^{-5} \nonumber \\
 & & (1-q^{13})^{-2} (1-q^\frac{40}{3})^{-15} (1-q^\frac{41}{3})^{-4}
 (1-q^{14})^{2} (1-q^\frac{43}{3})^{-4} \nonumber \\
 & & (1-q^\frac{44}{3})^{20} (1-q^{15})^{-2} (1-q^\frac{46}{3})^{14}
 (1-q^{16})^{-8} + \mathcal{O}(q^{\frac{50}{3}}) \,. 
 \label{efs_efc_Ising_tri}
\end{eqnarray}

As for the previous models defined on triangular lattices, FLM
calculations were also performed on lattices inscribed in
triangles. Having computed finite-lattice partition functions for
triangles of size $M \leq 23$, which yield series expansion up to
order $x^{84}$, that is $q^{\frac{42}{3}}$, we find for $\rme^{f_{\rm
    b}}$ the same expression as (\ref{product_efb_ising_tri}).
The first terms of $\rme^{f_{\rm s}}$ agree with
(\ref{efs_efc_Ising_tri}), except for the fact that all exponents are
here multiplied by $\frac{3}{2}$ due to the different definitions
(\ref{fMN_decomposition}) and (\ref{fM_decomposition}) chosen for
$f_{\rm s}$. Concerning the corner free energy, we find in this case
\begin{eqnarray}
 {\rm e}^{f_{\rm c}} &=& 2 (1-q^\frac{2}{3})^{-3} (1-q^\frac{8}{3})^{6}
 (1-q^\frac{10}{3})^{3} (1-q^\frac{12}{3})^{-1} (1-q^\frac{14}{3})^{-9}  \nonumber \\ & & (1-q^\frac{16}{3})^{-9} (1 - q^\frac{20}{3})^{9} (1 - q^\frac{22}{3})^9 (1-q^\frac{26}{3})^{-15} (1-q^\frac{28}{3})^{-9} \nonumber \\ 
 & & (1-q^\frac{32}{3})^{21} (1-q^\frac{34}{3})^{15} (1-q^\frac{36}{3})^{-1} (1-q^\frac{38}{3})^{-21} (1-q^\frac{40}{3})^{-24}
 \nonumber \\
 & & + \mathcal{O}(q^{\frac{42}{3}}) \,.
 \label{corner_ising_tri_tri}
\end{eqnarray}
To fix the corresponding factorised form, we first note that
eqs.~(\ref{efb_Isingsquare}) and (\ref{product_efb_ising_tri}) giving
${\rm e}^{f_{\rm b}}$ for the Ising model on respectively the square
and triangular lattices are both expressions of the general form
(\ref{generic factorized form}) with exponents $\beta_k$ and
$\gamma_k$ that are $8$-periodic in $q$. Seeing that ${\rm e}^{f_{\rm
    c}}$ for the square-lattice Ising model is also of this form, still with
periodicity $8$, we can assume the same to hold on the triangular
lattice.  The above development (\ref{corner_ising_tri_tri}) falls
slightly short of furnishing enough coefficients in order to fix
$\beta_k$ and $\gamma_k$ unambiguously. However, it is consistent with
the following appealing conjecture
\begin{eqnarray}
 {\rm e}^{f_{\rm c}} &=& 2 \prod_{k=1}^{\infty}{\frac{1}{1-q^{8k-4}}} \prod_{k=1}^{\infty}{\left(\frac{1-q^{8k-\frac{14}{3}}}{1-q^{8k-\frac{22}{3}}}\right)^{12k-9} \left(\frac{1-q^{8k-\frac{2}{3}}}{1-q^{8k-\frac{10}{3}}}\right)^{12k-3}} \nonumber  \\
& & \prod_{k=1}^{\infty}{\frac{\left(1-q^{8k-\frac{16}{3}}\right)^{15k-9}}{\left(1-q^{8k-\frac{8}{3}}\right)^{15k-6}}}  \prod_{k=1}^{\infty}{\frac{\left(1-q^{8k-\frac{4}{3}}\right)^{9k}}{\left(1-q^{8k-\frac{20}{3}}\right)^{9k-9}}} \,. 
\label{corner_ising_tri_tri_conj}
\end{eqnarray}
Our confidence that (\ref{corner_ising_tri_tri_conj}) is indeed
correct in enhanced by the fact that its asymptotic behaviour predicts
the same universal divergence of the correlation length on the square
and triangular lattices (see section~\ref{sec:ising_model_asymptotics}).

\subsubsection{Periodicities of product exponents.}
\label{sectionperiodicities}

Table~\ref{Periodicities} reviews the periodicities in terms of the
variable $q$ that we have observed in the exponents for the various
models described above.  It is seen that for the Potts and loop models
(where $q$ is the deformation parameter in the underlying quantum
group symmetry) the periodicities of the boundary-related quantities
${\rm e}^{f_{\rm s}}$ and ${\rm e}^{f_{\rm c}}$ is invariably twice
that of ${\rm e}^{f_{\rm b}}$.  On the other hand, for Ising models
(where $q$ enters the elliptic parameterisation of the coupling
constant) the periodicities of all three quantities is the
same. Another observation is than whenever the ``same'' model (Ising
or Potts) is solvable on two different lattices, the periodicities are
unchanged. It would be interesting to shed further light on these
observations, for instance from the perspective of the corner transfer
matrix and/or conformally invariant boundary states.

\begin{table}
\caption{\label{Periodicities} Periodicities of the exponents appearing in the factorized form of bulk, surface and corner free energies for different loop models.}
\begin{tabular}{lllll}
\br
Model & Lattice & ${\rm e}^{f_{\rm b}}$ periodicity & ${\rm e}^{f_{\rm s}}$ periodicity & ${\rm e}^{f_{\rm c}}$ periodicity \\
\mr
Potts selfdual  & Square     & 4 & 8 & 8 \\
Potts selfdual  & Triangular & 4 & 8 & 8 \\
Potts antiferromagnet & Square     & 8 & 16 & 16 \\
Chromatic polynomial  & Triangular & 6 & 12 & 12 \\
FPL${}^2$             & Square     & 8 & 16 & 16 \\
Ising                 & Square     & 8 &  8 &  8 \\
Ising                 & Triangular & 8 & 8 (?) & 8 (?) \\
\br
\end{tabular}
\end{table}
 
\subsection{Critical limits}
\label{critical limit 1}

In this section we study the critical limit(s) $q\to 1^{-}$ (and
$q\to -1^{+}$ whenever the latter corresponds to a critical theory) of
the product forms obtained for $\rme^{f_{\rm b}}$, $\rme^{f_{\rm s}}$
and $\rme^{f_{\rm c}}$. Physically, one expects a finite limit for
$\rme^{f_{\rm b}}$ and $\rme^{f_{\rm s}}$, whereas $\rme^{f_{\rm c}}$
is expected to exhibit a divergent behaviour at the critical point, which will be related
to a divergence of the characteristic dimensions of the system, such
as its correlation length $\xi$ \cite{CardyPeschel}.

We first discuss a few general properties observed when studying the
convergence of products having the structure (\ref{generic factorized
  form}), in the limit $q\to 1^{-}$. We present these properties
without formal proofs, but we demonstrate their applicability through
a series of concrete calculations that we exemplify below.

Let us first consider a product of the type (\ref{generic factorized
  form}) with purely periodical exponents $\alpha_k$ (that is, with
$\beta_k = 0$ for all $k$), as it is the case for most of the
expressions found in the course of this work. We can emphasize the
periodicity property of  such a product by rewriting it as
\begin{equation}
 \prod_{k=1}^{\infty} 
 \left(1-q^{ak-(a-1)}\right)^{\gamma_1}
 \left(1-q^{ak-(a-2)}\right)^{\gamma_2} \cdots
 \left(1-q^{ak}\right)^{\gamma_a} \,,
\end{equation}
where $a \in \mathbb{N}$ denotes the period.  We have proved that
this product has a finite limit when $q\to 1^{-}$ if and only if the
two following conditions are satisfied
\begin{eqnarray}
 \gamma_1 + \gamma_2 + \ldots +\gamma_{a} & = & 0 \,, \\
 \gamma_1 (a-1) + \gamma_2 (a-2) + \ldots +\gamma_a 0 &=& 0 \,.
\end{eqnarray}
If $\gamma_1 + \gamma_2 + \ldots +\gamma_{a} > 0$ (resp.\ $<0$), the
product is divergent (resp.\ has a zero limit). If $ \gamma_1 +
\gamma_2 + \ldots +\gamma_{a} = 0$, the product is divergent for
$\gamma_1 (a-1) + \gamma_2 (a-2) + \ldots +\gamma_a 0 < 0$ and has a
zero limit for $\gamma_1 (a-1) + \gamma_2 (a-2) + \ldots +\gamma_a 0 >
0$.

Now turn to the general form (\ref{generic factorized form}), where
$\alpha_k=\beta_k k + \gamma_k$. We have not found any solid
analytical arguments enabling the classification of the $q\to 1^{-}$
limit or the asymptotical behaviour of such products. However, the
following property, yet unexplained, will turn out to be useful in the
description of the Ising model corner free energy divergence: it was
observed that products of the type
\begin{equation}
 \prod_{k=1}^{\infty}{\frac{(1-q^{ak-m})^{ak-m}}{(1-q^{ak-n})^{ak-n}}}
\end{equation}
have a finite limit when $q\to 1^-$ if and only if $m+n = a$ (not
mentionning the trivial case $m=n$, for which the product equals $1$
for any $q$).

\subsubsection{Finite limits of bulk and surface free energies.}

To evaluate the limits of the $\rme^{f_{\rm b}}$ and $\rme^{f_{\rm
    s}}$ products, each factor $(1-q^l)^{\alpha_l}$ can be replaced by
$(l(1-q))^{\alpha_l}$. Using the identity
\begin{equation}
 \Gamma(z) = -z^{-1} \prod_{k=1}^{\infty}
 {\left[\left(1+\frac{1}{k}\right)^z \left(1-\frac{z}{k}\right)^{-1}\right]} \,,
\end{equation}
where $\Gamma(z)$ is the Euler gamma function, one is led to the
following results.

\begin{itemize}

\item For the selfdual Potts model on the square lattice:
\begin{eqnarray}
 \lim_{q\to 1^{-}} {\rm e}^{f_{\rm b}} &=&
 18 \frac{\Gamma\left(-\frac{3}{4}\right)^2
 \Gamma\left(\frac{1}{4}\right)^2}{\Gamma\left(-\frac{1}{4}\right)^4} \,, \\
 \lim_{q\to 1^{-}} {\rm e}^{f_{\rm s}} &=&
 8 \frac{\Gamma\left(\frac{3}{8}\right)^2}
 {\Gamma\left(-\frac{1}{8}\right)^2} \,.
\end{eqnarray}

\item For the critical antiferromagnetic Potts model on the square lattice:
\begin{eqnarray}
 \lim_{q\to 1^{-}} {\rm e}^{f_{\rm b}} &=& 0 \,, \\
 \lim_{q\to 1^{-}} {\rm e}^{f_{\rm s}} &=&
 -\frac{5}{3}\frac{\Gamma\left(\frac{3}{8}\right)
 \Gamma\left(\frac{1}{8}\right)}{\Gamma\left(\frac{-1}{8}
 \Gamma\left(\frac{5}{8}\right)\right)} \,.
\end{eqnarray}
The $q\to 1^-$ (that is, $Q\to 0^-$) limit of the bulk free energy is
trivial, since in this case $Z=0$ both in finite-size and in the
thermodynamic limit. However, dividing the finite-lattice partition
functions $Z_{MN}$ by $Q^{M N}$ produces a model which in the $Q \to
0$ limit has a non-trivial combinatorial interpretation
\cite{forests1,forests2}.  It amounts to counting so-called
\textit{forests}, which are collections of spanning trees (i.e.,
Fortuin-Kasteleyn clusters without cycles), on the square lattice,
where each component tree carries the fugacity $w=-4$. The
corresponding partition function is given by
\begin{equation}
 (Z_{\rm forests})^{1/(MN)} =
 \lim_{q\to 1^{-}} \frac{{\rm e}^{f_{\rm b}}}{Q} =
 \frac{\Gamma\left(\frac{5}{4}\right)^2}{\Gamma\left(\frac{3}{4}\right)^2} \,.
\end{equation}
Not much more about this problem seems to be computable from the
expression for ${\rm e}^{f_{\rm b}}$, since its derivatives with
respect to $Q$ involve clusters with cycles. In particular, it is
unfortunately not possible to compute the average number of trees by
this method.

\item For the selfdual Potts model on the triangular lattice:
\begin{eqnarray}
 \lim_{q\to 1^{-}} {\rm e}^{f_{\rm b}} &=&
 \frac{108}{\pi^{3/2}}\frac{\Gamma\left(\frac{7}{6}\right)^6}
 {\Gamma\left(\frac{5}{4}\right)^2\Gamma\left(\frac{5}{6}\right)^3} \,, \\
 \lim_{q\to 1^{-}} {\rm e}^{f_{\rm s}} &=& 4-2\sqrt{3} \,.
\end{eqnarray}

\item For the chromatic polynomial on the triangular lattice
  \cite{JacobsenChromatic}:
\begin{eqnarray}
 \lim_{q\to 1^{-}} {\rm e}^{f_{\rm b}} &=&
 -54 \times 2^{1/3} \frac{\Gamma\left(\frac{7}{6}\right)^4 }
 {\pi^2\Gamma\left(\frac{2}{3}\right)} \,, \\
 \lim_{q\to 1^{-}} {\rm e}^{f_{\rm s}} &=&
 8 \frac{\Gamma\left(\frac{5}{12}\right)^2
 \Gamma\left(\frac{5}{4}\right)^2}{\pi\Gamma\left(\frac{1}{6}\right)^2} \,.
\end{eqnarray}

\item For the FPL${}^2$ model:
\begin{eqnarray}
 \lim_{q\to 1^{-}} {\rm e}^{f_{\rm b}} &=&
 \frac{\Gamma\left(\frac{1}{4}\right)^8}{4\pi^6} \,, \\
 \lim_{q\to 1^{-}} {\rm e}^{f_{\rm s}} &=&
 144 \sqrt{2} \pi \frac{\Gamma\left(-\frac{3}{4}\right)^2}
 {\Gamma\left(-\frac{1}{8}\right)^4} \,.
\end{eqnarray}

\end{itemize}

\subsubsection{Corner free energy divergence.}

Using the previous method to compute the limit of ${\rm e}^{f_{\rm c}}$,
one finds that the corresponding factorised form is either divergent
or has a zero limit. We shall see below that these two possibilities
distinguish the sign of the central charge $c$ of the CFT that emerges
in the $q \to 1^-$ (or $q \to -1^+$) limit. A more detailed study of
the asymptotics of ${\rm e}^{f_{\rm c}}$ will enable us to determine the
precise value of $c$ as well as the precise asymptotic divergence of
the correlation length $\xi$ in the critical limit.

In this section we first expose the main tools needed in the
asymptotic analysis, and we give precise results for each of the
models for which we have been able to obtain an exact product formula
for ${\rm e}^{f_{\rm c}}$.  In the following section
\ref{sec:Cardy_Peschel} we confront these results with the CFT
predictions \cite{CardyPeschel}.

To obtain the asymptotical behaviour of infinite products of the type
(\ref{generic factorized form}), in the limit $q\to 1^{-}$, one can
use the properties of the Dedekind eta function, defined in the upper
half of the complex plane by
\begin{equation}
\eta(\tau)=q^{\frac{1}{24}}\prod_{k=1}^{\infty}{(1-q^k)} \,,
\end{equation}
where $q\equiv \rme^{2\mathrm{i} \pi \tau}$.  Setting $\tau=\mathrm{i}
x$, and using the modular transformation identity \cite{Dedekind}
\begin{equation}
\eta(-\tau^{-1})=\sqrt{-\mathrm{i} \tau}\eta(\tau) \,,
\end{equation}
one can deduce the asymptotical behaviour of $\eta(\mathrm{i} x)$ as
$q\to1^{-}$ (resp.\ $x\to 0^{+}$) from its Taylor expansion as $q\to 0$
(resp.\ $x\to +\infty$). Finally, we find the general formula
\begin{equation}
 \prod_{k=1}^{\infty}{\frac{1}{1-q^k}} \stackrel[q\to 1^{-}]{}{\sim}
 \sqrt{\frac{-\log q}{2\pi}} \rme^{\frac{-\pi^2}{6 \log q}} \,.
\end{equation} 
In practice, we will use the related formula
\begin{equation}
 \prod_{k=1}^{\infty}{\frac{1}{1-q^{\alpha k}}} \stackrel[q\to 1^{-}]{}{\sim}
 \sqrt{\alpha \frac{1-q}{2\pi}}
 {\rm e}^{\frac{\pi^2}{6 \alpha(1-q)}-\frac{\pi^2}{12\alpha}} \,,
\end{equation} 
from which the asymptotics of the different product forms obtained for
${\rm e}^{f_{\rm c}}$ can easily be obtained as follows.

\begin{itemize}
\item For the selfdual Potts model on the square lattice one finds
\begin{eqnarray}
 {\rm e}^{f_{\rm c}} &=& \prod_{k=1}^{\infty}
 \frac{1}{(1-q^{8k-6}) (1-q^{8k-4})^4 (1-q^{8k-2})} \nonumber \\
 &=& \left( \prod_{k=1}^{\infty} \frac{1}{1-(q^2)^{k}} \right)
 \left( \prod_{k=1}^{\infty} \frac{1}{1-(q^4)^{k}} \right)^3
 \left( \prod_{k=1}^{\infty} \frac{1}{1-(q^8)^{k}} \right)^{-4} \nonumber \\
 & \stackrel[q\to 1^{-}]{}{\sim} &
 2^{-5/2} {\rm e}^{\frac{\pi^2}{8} \left(\frac{1}{1-q}-\frac{1}{2}\right)} \,.
\end{eqnarray}

\item For the antiferromagnetic critical Potts model on the square lattice, we find by
  similar considerations
\begin{equation}
 {\rm e}^{f_{\rm c}} \stackrel[q\to 1^{-}]{}{\sim}
 {\rm e}^{-\frac{\pi^2}{16} \left(\frac{1}{1-q}-\frac{1}{2}\right)} \,.
\label{eq:divergence antiferro} 
\end{equation}

\item For the selfdual Potts model on the triangular lattice the
  divergence of the corner free energy (\ref{efc_Potts_tri_tri}), that
  is, the corner free energy of the three corners of angle
  $\frac{\pi}{3}$ in a triangle-shaped lattice, is found to be
\begin{equation}
 {\rm e}^{f_{\rm c}} \stackrel[q\to 1^{-}]{}{\sim} \frac{3\sqrt{3}-5}{2}
 {\rm e}^{\frac{\pi^2}{6} \left(\frac{1}{1-q}-\frac{1}{2}\right)} \,.
\label{eq:divergence_Potts_tri_tri}
\end{equation}

\item For the chromatic polynomial on the triangular lattice, the
  limit $x\to -1^+$ of the corner free energy (that is, $q\to 1^{-}$
  or $Q\to 4^+$) can be computed by turning the product (\ref{efc
    chromatic rectangle}) into a function of $q$, as indicated in
  (\ref{x to q}). This however leads to a large number of factors and
  lengthy calculations. We therefore indicate here an alternative way
  of evaluating the behaviour of the divergent contributions directly
  in terms of the variable $x$.  In the product expression (\ref{efc
    chromatic rectangle}) for $\rme^{f_{\rm c}}$ the factors with an
  even power of $x$ yield a finite contribution in the limit $x\to
  -1^+$, whereas the diverging contribution is given by the factors
  with odd powers of $x$, namely of the type
  $1-x^{12k-(2p+1)}$. Series expanding the logarithm of these factors,
  permuting summations, and summing the geometric series over $k$ we
  find
\begin{eqnarray}
 \log \prod_{k=1}^{\infty}\left(1-x^{12k-2p-1}\right) &=&
 -\sum_{k=1}^{\infty}\sum_{n=1}^{\infty}\frac{x^{(12k-2p-1)n}}{n} \nonumber \\
 &=& -\sum_{n=1}^{\infty}{\frac{x^{-(2p+1)n}}{n}\frac{1}{1-x^{12n}}} \nonumber \\
 &\stackrel[x\to -1^{+}]{}{\sim}&  
 -\sum_{n=1}^{\infty}{\frac{x^{-(2p+1)n}}{12n^2(1+x)}} \nonumber \\
 &\stackrel[x\to -1^{+}]{}{\sim}& \frac{\pi^2}{144(1+x)} \,,
 \end{eqnarray}
 that is, in terms of $q$,
 \begin{equation}
 \log \prod_{k=1}^{\infty}\left(1-(-q^2)^{12k-2p-1}\right) \stackrel[q\to 1^{-}]{}{\sim}  \frac{\pi^2}{288(1-q)} \,, 
\end{equation}
independently of $p$.  Summing up all the contributions, we find for the energies
$f_{\rm c }\left(\frac{\pi}{3}\right)$ associated with corners of angles $\frac{\pi}{3}$ and $\frac{2\pi}{3}$ respectively the following behaviours
\begin{eqnarray}
 f_{\rm c}\left(\frac{\pi}{3}\right)  & \stackrel[q\to 1^{-}]{}{\sim} &
 \frac{7\pi^2}{432}\frac{1}{1-q} \,,
 \label{eq:divergence fc chromatic1} \\
  f_{\rm c}\left(\frac{2\pi}{3}\right)  & \stackrel[q\to 1^{-}]{}{\sim} &
 \frac{7\pi^2}{864}\frac{1}{1-q}  \,.
\end{eqnarray}
The limit $x\to 1^-$ (that is, $Q\to 0^-$) can be computed directly in
terms of $x$, and we find using the same methods as for other models
\begin{eqnarray}
 f_{\rm c}\left(\frac{\pi}{3}\right)  & \stackrel[x\to 1^{-}]{}{\sim} &
 -\frac{7\pi^2}{108}\frac{1}{1-x} \,, \\
  f_{\rm c}\left(\frac{2\pi}{3}\right)  & \stackrel[x\to 1^{-}]{}{\sim} &
 -\frac{7\pi^2}{216}\frac{1}{1-x}  \,.
 \label{eq:divergence fc chromatic2} 
\end{eqnarray}

\item For the FPL${}^2$ model 
\begin{equation}
 {\rm e}^{f_{\rm c}} \stackrel[q\to 1^{-}]{}{\sim}
 \frac{1}{4} {\rm e}^{\frac{3\pi^2}{16} \left(\frac{1}{1-q}-\frac{1}{2}\right)} \,.
\label{eq:divergence FPL} 
\end{equation}

\item For the square-lattice Ising model, the properties stated at the
  beginning of section \ref{critical limit 1} allow us to isolate the
  divergent contribution of the exponentiated corner free energy from
  finite factors. We thus find
\begin{eqnarray}
 {\rm e}^{f_{\rm c}} &\stackrel[q\to 1^{-}]{}{\propto}&
 \prod_{k=1}^{\infty}{\frac{(1-q^{8k-6})^3(1-q^{8k-2})^3}{(1-q^{8k-4})(1-q^{2k-1})^2}}
 \nonumber \\
 &\stackrel[q\to 1^{-}]{}{\sim}& {\rm e}^{\frac{\pi^2}{16(1-q)}} \,. 
\end{eqnarray}

\item For the triangular-lattice Ising model inscribed in an
  equilateral triangle, using similar methods, 
  the divergence of $\rme^{f_{\rm c}}$ in
  (\ref{corner_ising_tri_tri_conj}) is found to be
\begin{equation}
 {\rm e}^{f_{\rm c}} \stackrel[q\to 1^{-}]{}{\propto} {\rm e}^{\frac{\pi^2}{12(1-q)}} \,. 
 \label{divergence_efc_ising_tri_tri}
\end{equation}

\end{itemize}

\subsection{Relation to conformal field theory}
\label{sec:Cardy_Peschel}

Our interest here is to understand these corner free energy
divergences by the standards of conformal field theory (CFT). Cardy
and Peschel showed in \cite{CardyPeschel} that the presence of a
corner of interior angle $\gamma$ along boundaries of typical size $L$
in a two-dimensional conformally invariant (hence, critical) model of
central charge $c$ originates from a logarithmic correction to the
free energy, namely
\begin{eqnarray}
 \Delta F &=& -\frac{c\gamma}{24\pi}\left(1-(\pi/\gamma)^2\right)\ln L
 \nonumber \\
 &=& \left \lbrace
 \begin{array}{ll}
 \frac{c}{16} \ln L    & \mbox{for $\gamma=\pi/2$} \,, \\
 \frac{c}{9} \ln L     & \mbox{for $\gamma=\pi/3$} \,, \\
 \frac{5c}{144} \ln L  & \mbox{for $\gamma=2\pi/3$} \,. \\
 \end{array}
 \right.
\label{eq:cardy}
\end{eqnarray}
Note the different sign compared to Cardy and Peschel's explicit
formula, due to the fact that we chose to define the free energy as
the logarithm of the partition function, without the conventional
minus sign.

We now wish to compare the results obtained above with this
prediction. We emphasise the fact that the FLM formulae themselves
forbid the appearance of $\ln L$ like terms, since all the
contributions $f_{\rm b}$, $f_{\rm s}$ and $f_{\rm c}$ are by
construction independent of the system's dimensions.  Indeed, at the
critical point $q=1$, the series expansion in powers of $q$ breaks
down, since all the finite subgraphs so far neglected by the FLM
formulae become non-negligible.  This problem can be reformulated in
other terms: if the $M\times N$ lattice is considered finite, Enting's
formulae give an exact result up to order $k \sim M+N \equiv L$. When
approaching criticality ($q\to 1^-$), the cutoff $B(k)$ has thus to be
chosen bigger and bigger in order to reach a given
precision in the series approximation, and so must therefore the size
of the system. So to summarise, our use of the FLM formulae is valid
at criticality only provided that the size $L$ of the system diverges
fast enough to include excitations of increasing size, that is,
diverges as the correlation length. In this context
eq.~(\ref{eq:cardy}) can thus be rewritten as
\begin{equation}
 \Delta F = -\frac{c\gamma}{24\pi}\left(1-(\pi/\gamma)^2\right)\ln \xi \,,
\label{eq:cardy2}
\end{equation}
where $\xi(q)$ is some characteristic length defining the typical
maximal size of excitations that enter the FLM formulae with a
non-negligible statistical weight.

\subsubsection{Selfdual Potts model.}

For the selfdual Potts model on the {\em square} lattice, we have $c=1$ when
approaching the critical region via the limit $q\to1^-$.  Recalling
that the corner free energy divergence is the sum of four
$\gamma=\pi/2$ contributions of the type (\ref{eq:cardy}), we thus
have
\begin{equation}
 \ln \xi \stackrel[q\to 1^{-}]{}{\sim} \frac{\pi^2}{2(1-q)} \,,
\end{equation}
or more precisely,
\begin{equation}
 \xi \stackrel[q\to 1^{-}]{}{\sim}
 \frac{1}{2^{10} {\rm e}^{\pi^2/4}} \ {\rm e}^{\frac{\pi^2}{2(1-q)}} \,.
 \label{sq_Potts_xi}
\end{equation}

It is of course tempting (as suggested by our notation) to interpret
$\xi$ as a correlation length for the near-critical system. We shall
now see that this interpretation is indeed correct. Based on Bethe
Ansatz (BA) calculations, Wallon and Buffenoir \cite{Wallon} have
identified the correlation length $\xi_{\rm BA}$ from the ratio of the
leading and next-leading eigenvalues of the XXZ spin chain
hamiltonian. They obtain asymptotically
\begin{equation}
 \xi_{\rm BA} \stackrel[q\to 1^{-}]{}{\sim}
 \frac{1}{8\sqrt{2}} \ {\rm e}^{\frac{\pi^2}{2(1-q)}} \,.
\end{equation}
This coincides with (\ref{sq_Potts_xi}) up to a $q$-independent
multiplicative prefactor (of numerical value $\simeq 1068$).
Crucially, the constant $\frac{\pi^2}{2}$ in the exponential,
governing the strength of the essential singularity, is precisely the
same. The multiplicative prefactor was of course to be expected, since
a system does not have just one correlation length, but several, which
are all proportional and thus present the same critical divergence.
We can therefore safely refer to $\xi$ as a (and sometimes, by an
abuse of language, even ``the'') correlation length.

We now turn to the selfdual Potts model on the {\em triangular} lattice,
for which eq.~(\ref{eq:divergence_Potts_tri_tri}) provides the asymptotics
of the corner free energy for three corners of angle $\frac{\pi}{3}$.
In conjunction with eqs.~(\ref{eq:cardy})--(\ref{eq:cardy2}), and using
that $c=1$ is lattice-independent, we obtain
\begin{equation}
 \xi \stackrel[q\to 1^{-}]{}{\sim}
 \frac{(3\sqrt{3}-5)^3}{8 {\rm e}^{\pi^2/4}} \ {\rm e}^{\frac{\pi^2}{2(1-q)}} \,.
 \label{tri_Potts_xi}
\end{equation}
The fact that the universal, $q$-dependent part of the asymptotics
precisely coincides with (\ref{sq_Potts_xi}) for the square lattice
is a nice verification of universality, and of the angular dependence 
appearing in the Cardy-Peschel formula (\ref{eq:cardy}).

Note also that ratio of the correlation lengths on the two lattices
tends asymptotically to a constant,
\begin{equation}
 \lim{q\to 1^-} \frac{\xi_{\rm tri}}{\xi_{\rm sq}} = 128 (3 \sqrt{3}-5)^3 \simeq 0.966031.
\end{equation}
It would be interesting to confront this prediction with numerical
results, such as Monte-Carlo simulations.

In section \ref{sec:boundaries} below we shall investigate how the
effective central charge---extracted from the divergence of the corner
free energy using (\ref{eq:cardy})---is affected by changing the
boundary conditions at the position of the corner. This change can be intrepreted
within CFT as the insertion of a boundary condition changing operator in
the corner.

\subsubsection{Antiferromagnetic Potts model.}

For the antiferromagnetic Potts model on the {\em square} lattice, the
limit $q \to 1^-$ (or $Q \to 0^-$) corresponds to a critical theory
with $c=-1$
\cite{Saleur91,JacobsenSaleurAntiferro,IkhlefJacobsenSaleur}.  The
comparison between (\ref{eq:divergence antiferro}) and
(\ref{eq:cardy2}) gives the following prediction for the behaviour of
the correlation length
\begin{equation}
 \xi(q) \stackrel[q\to 1^{-}]{}{\sim}
 {\rm e}^{\frac{\pi^2}{4}\left(\frac{1}{1-q}-\frac{1}{2}\right)} \,.
 \label{xi_AF}
\end{equation}

As already mentioned in section \ref{sec:triangular-lattice_Potts},
numerical evidence has been given \cite{JacobsenSaleurAntiferro} that
the same $c=-1$ CFT arises as the continuum limit of the {\em
  triangular}-lattice selfdual Potts model, in the limit where $t \to
-1^+$, i.e., $(Q,v) \to (0,-3)$.
The corresponding correlation length can be computed
from (\ref{efc_Potts_tri_tri}), which is shown to diverge
for $t\to -1^+$ like
\begin{equation}
 \rme^{f_{\rm c}} \stackrel[t\to -1^{+}]{}{\sim}
 \rme^{-\frac{\pi^2}{18}\left(\frac{1}{1+t}-\frac{1}{2}\right)} \,. 
\end{equation}
Comparison with (\ref{eq:cardy2}) yields for the associated correlation length
\begin{equation}
 \xi(t) \stackrel[t\to -1^{+}]{}{\sim} 
 {\rm e}^{\frac{\pi^2}{6} \left( \frac{1}{1+t} - \frac{1}{2} \right)} \,.
 \label{xi_tri_-1}
\end{equation}

In order to compare the correlation lengths (\ref{xi_AF}) and
(\ref{xi_tri_-1}), we need to express both in terms of the same
parameter, most naturally $Q$. We find from the respective definitions of
$q$ and $t$ that
\begin{equation}
 2(1-q) \stackrel[Q\to 0^{-}]{}{=}
 \sqrt{-Q}\left(1-\frac{1}{4}\sqrt{-Q} + \mathcal{O}(Q)\right) 
\end{equation}
for the antiferromagnetic square-lattice Potts model, and 
\begin{equation}
 3(1+t) \stackrel[Q\to 0^{-}]{}{=}
 \sqrt{-Q}\left(1-\frac{1}{6}\sqrt{-Q} + \mathcal{O}(Q)\right) 
\end{equation} 
for the triangular-lattice selfdual Potts model. Both results can
thus be rewritten in the common form
\begin{equation}
 \xi(t) \stackrel[Q\to 0^{-}]{}{\sim} {\rm e}^{\frac{\pi^2}{2\sqrt{-Q}}} \,.
 \label{xi_tri_-1bis}
\end{equation}
This provides a verification of the universality between the two models, which
goes beyond the numerical evidence of \cite{JacobsenSaleurAntiferro}. Note that
not only do the two correlation lengths given above have the same critical divergence,
but the prefactors are also identical (and equal to unity).

\subsubsection{Chromatic polynomial.}

For the chromatic polynomial we can perturbatively access two
different critical theories. In the limit $q \to 1^-$ (or $Q \to 4^+$)
we have a CFT with $c=2$, whereas the limit $q \to i \times 1^-$ (or
$Q \to 0^-$, or $x \to 1^-$) one finds $c=-2$ (see \cite{JacobsenSalasSokal} and
references therein).  In both cases we must pay attention to the fact
that the corner angles in (\ref{eq:cardy2}) have to be adapted to the
triangular geometry.

For the $q \to 1^-$ limit we find
\begin{equation}
  \ln \xi(q) \stackrel[q\to 1^{-}]{}{\sim} \frac{7\pi^2}{96(1-q)} 
\end{equation}
for a corner of angle $\frac{\pi}{3}$, and
\begin{equation}
  \ln \xi(q) \stackrel[q\to 1^{-}]{}{\sim} \frac{7\pi^2}{30(1-q)} \,.
\end{equation}
for a corner of angle $\frac{2\pi}{3}$.
Meanwhile, for the $q \to i \times 1^-$ limit we find as a function of $x$ 
\begin{equation}
 \ln \xi(q) \stackrel[x\to -1^{-}]{}{\sim}
 \frac{7\pi^2}{24(1-x)} \,.
\end{equation}
for a corner of angle $\frac{\pi}{3}$, and
\begin{equation}
  \ln \xi(q) \stackrel[x\to -1^{-}]{}{\sim} \frac{7\pi^2}{15(1-x)} \,.
\end{equation}
for a corner of angle $\frac{2\pi}{3}$.

It is obviously worrying that for both limits the results for the
divergence of $\ln \xi(q)$ depend on the corner angle, i.e., are
mutually inconsistent.  On the other hand, we have seen that both the
results for the selfdual Potts model and those for the antiferromagnetic
Potts model nicely confirm universality
between the square and triangular lattices, once the angular
dependence of (\ref{eq:cardy}) has been taken into account. (We shall
also obtain a similar agreement for the Ising model in
section~\ref{sec:ising_model_asymptotics} below.)

One possible explanation for the discrepancy is that the corner
angles seen in the continuum limit do not equal those on the
lattice (i.e., $\frac{\pi}{3}$ and $\frac{2\pi}{3}$). This might
be the case for geometrically highly constrained (frustrated)
problems, such as the $4$-colourings under consideration.

An example of a situation where such a scenario is known to hold true
is provided by dimer coverings of the so-called Aztec diamond lattice,
which is just a diagonally oriented square piece of the square
lattice. It has been proved rigorously \cite{Aztec} that there are $Z =
2^{N(N+1)/2}$ dimer coverings of an Aztec diamond of size $N$. It
follows that $f_{\rm c} = 0$ exactly for any $N$.  Meanwhile, it is
well-known that dimer coverings of the square lattice are in bijection
with configurations of a scalar height, whose continuum limit is a free
Gaussian field. Therefore the central charge is $c=1$. This would seem at
odds with the result $f_{\rm c}=0$, since the Cardy-Peschel result
(\ref{eq:cardy}) then predicts that $f_{\rm c}$ depends non-trivially on
$N$. The resolution of the apparent paradox is that, due to the
boundary conditions, the dynamics of dimers close to the corners of
the lattice is frozen. In the thermodynamic limit, the region where
dimers are free to move becomes---by the so-called Arctic circle
theorem \cite{arctic}---a circle inscribed in the Aztec diamond. In this
sense there are {\em no corners} in the continuum limit, and the paradox
is resolved.

We believe that it would be interesting to investigate whether the
$4$-colouring problem has frozen dynamics close to the corners in the
thermodynamic limit.

\subsubsection{FPL${}^2$ model.}

The FPL$^2$ model has central charge $c=3$ when $q\to 1^-$
\cite{Kondev,JacobsenKondev}. We thus find from (\ref{eq:divergence
  FPL}) that
\begin{equation}
 \ln \xi(q) \stackrel[q\to 1^{-}]{}{\sim} \frac{\pi^2}{4(1-q)} \,.
\end{equation} 

\subsubsection{Ising model.}
\label{sec:ising_model_asymptotics}

\begin{itemize}
 \item For the Ising model on the square lattice, of central charge
$c=\frac{1}{2}$ at criticality, we find
\begin{equation}
 \ln \xi(q) \stackrel[q\to 1^{-}]{}{\sim} \frac{\pi^2}{2(1-q)} \,.
\label{asymptotics_xi_ising_q}
\end{equation}

This result can be compared to the analytical expression given in
\cite{BaxterBook} for the Ising model correlation length in the
critical limit, namely
\begin{equation}
\xi^{-1} \sim -\ln k \,,
\label{Baxter_xi_Ising}
\end{equation}
where $k$ is defined as
\begin{equation}
k=(\sinh 2K)^{-2} \,.
\label{Baxter_k_Ising}
\end{equation}
Rewriting this as a function of the temperature $T$, we have
\begin{equation}
\xi \stackrel[T\to T_{\rm c}]{}{\propto} \frac{1}{|T-T_{\rm c}|} \,,
\end{equation}
from which one infers the value $\nu = 1$ of the correlation length
critical exponent. In order to retrieve this well-known result from
(\ref{asymptotics_xi_ising_q}), we need to re-express the latter in
terms of $T$.

It is easily checked that below the critical temperature one has
\begin{equation}
 x_{\rm c}-x \stackrel[T\to {T_{\rm c}}^{-}]{}{\propto} T_{\rm c} - T \,.
\end{equation}
We thus need to compute the asymptotical behaviour of $x$ in terms of
$q$ when $q\to 1^{-}$.  This is done in terms of elliptic functions
of modulus $k$ from eq.~(3) of \cite{BaxterIsing}, namely (with the
usual notations)
\begin{equation}
 x^2 = -\mathrm{i} k^{1/2} \mathrm{sn} \frac{\mathrm{i}K'}{4} \,,
 \label{x2_Ising_elliptic}
\end{equation}
where $q=\exp\left(-\frac{K'}{4K}\right)$. The limit $q\to 1^{-}$ is
obtained for $k\to 1^{-}$, yielding for $K$ and $K'$ the following
expansions
\begin{eqnarray}
 K & \sim & -\frac{1}{2}\ln(1-k^2)\left(1-\frac{k^2-1}{4} \right) + \ln
 4 + \mathcal{O}\left(1-k^2\right) \,, \\
 K' & \sim & \frac{\pi}{2}\left(1+\frac{k^2-1}{4} +
 \frac{9}{64}\left(1-k^2\right)^2\right) +
 \mathcal{O}\left((1-k^2)^3\right) \,.
 \label{series_elliptic}
\end{eqnarray}
Series expanding eq.~(\ref{x2_Ising_elliptic}) around $k^2 \equiv m \to 1^-$ yields
\begin{equation}
 x^2 = -1+\sqrt{2} -8\left(2-\sqrt{2}\right) (1-k^2)^2
 +\mathcal{O}\left((1-k^2)^3\right) \,,
\end{equation}
that is, to leading order,
\begin{equation}
 x_{\rm c} - x \stackrel[k\to 1^-]{}{\propto} {x_{\rm c}}^2 - x^2 \stackrel[k\to
 1^-]{}{\propto} -(1-k^2) \,.
\end{equation}
{}From the above definition of $q$ in terms of $K$ and $K'$ we furthermore have
\begin{equation}
 1-q \stackrel[k\to 1^-]{}{\sim } -\frac{\pi^2}{4 \ln(1-k)} \,,
\end{equation}
and thus,
\begin{equation}
  T_{\rm c} - T \stackrel[T\to {T_{\rm c}}^-]{}{\sim}
  (1-k)^2 \stackrel[q\to 1^-]{}{\sim}
  \rme^{-\frac{\pi^2}{2(1-q)}} \,.
 \end{equation}
We can thus rewrite the divergence (\ref{asymptotics_xi_ising_q}) of
the correlation length in terms of $T_{\rm c}-T$, which yields, as expected,
\begin {equation}
 \xi \stackrel[T\to {T_{\rm c}}^-]{}{\propto } \frac{1}{|T-T_{\rm c}|} \,.
 \label{xi_T_ising_sq}
\end{equation}%

\item For the triangular-lattice Ising model, in the case of a lattice
  inscribed in an equilateral triangle, we find from
  (\ref{divergence_efc_ising_tri_tri}) that
\begin{equation}
 \ln \xi(q) \stackrel[q\to 1^{-}]{}{\sim} \frac{\pi^2}{2(1-q)} \,.
\label{asymptotics_xi_ising_q_bis}
\end{equation}
In the same way as for the square lattice, we seek an expression of the parameter $x$ defined in (\ref{param_x_Ising_tri}) in terms of elliptic functions. One possibility is given by
\begin{equation}
x^2 = -\mathrm{i}k^{\frac{1}{2}} \mathrm{sn}\left(\frac{\mathrm{i}K'}{3}\right) \,,
\end{equation}
where, once again, $q=\exp\left(-\frac{K'}{4K}\right)$.  Using
(\ref{series_elliptic}), we find that around the critical point
${x_{\rm c}}^2 = \frac{1}{\sqrt{3}}$,
\begin{equation}
x^2-{x_{\rm c}}^2 = -\frac{(1-k)^2}{32\sqrt{3}} \,,
\end{equation}
that is, as in the square-lattice case, 
\begin{equation}
T_{\rm c} - T \stackrel[T\to {T_{\rm c}}^-]{}{\sim } (1-k)^2 \,.
\end{equation}
We thus have once again
 \begin {equation}
 \xi \stackrel[T\to {T_{\rm c}}^-]{}{\propto } \frac{1}{|T-T_{\rm c}|} \,,
 \label{}
 \end{equation}
which gives in comparison with (\ref{xi_T_ising_sq}) a verification of
universality between square and triangular-lattice Ising models.
 
\end{itemize}

\section{Particular boundary conditions}
\label{sec:boundaries}

We now consider the effect of particular (non-free) boundary
conditions on the previous models. In section \ref{FLMbound}, we
derive FLM formulae for the case of lattices where one or more sides
are endowed with particular boundary conditions.  These formulae allow
us to compute the surface free energy with the particular boundary
conditions, or the corner free energy for a corner between two sides
with free-particular or particular-particular boundary conditions.
The resulting expressions will be compared with those found in section
\ref{sec:free_bcs} for the case of free boundaries boundary
conditions.

As will be discussed in further detail in section~\ref{FLMboundIsing}, the
Ising model presents some special difficulties, whose resolution we
leave for future work.  Therefore, we will focus for the remainder of
this paper on the selfdual Potts model on the square lattice.%
\footnote{We could equallly well have studied any of the other Potts
  or fully-packed loop models with particular boundary conditions.
  But technically it is easier to tackle the square-lattice Potts
  model, since in this case 1) the coefficients of the relevant
  product formulae present the smallest periodicity, and 2) our
  numerical methods for computing finite-lattice free energies are
  more powerful.}  The particular boundary conditions that we shall be
interested in are those introduced by Jacobsen and Saleur
\cite{JacobsenSaleurConformalBoundary}.  We refer to them as JS
boundary conditions for short, and define them precisely in
section~\ref{Parametrization}.

Our results for JS boundary conditions are presented in section
\ref{Resultsbound}. The critical limit $q\to 1^-$ is studied in
section~\ref{Criticalbound}, where we also make contact with the CFT results
of \cite{JacobsenSaleurConformalBoundary}. In particular, we shall see
how physical observables such as correlation length and conformal
properties are modified by the change of boundary conditions.

\subsection{FLM for particular boundary conditions}
\label{FLMbound}

In the case of lattices with particular boundary conditions on one or
several sides, the finite-lattice expansions given in section
\ref{sec:free_bcs} must be modified to take into account new types of
contributions, where the $[i,j]$ sublattice touches one or more sides
of the $[M,N]$ lattice.

In \cite{Enting2}, Enting has shown how to derive numerically FLM
formulae in the case of fixed boundary conditions for the 3-state
Potts model. In this section, we demonstrate how particular boundary
conditions can be dealt with analytically in a framework that does not
depend on the explicit choice of model or boundary conditions.  In
particular, we shall derive explicit formulae for the different
configurations corresponding to one, two, three our four sides with
particular boundary conditions, regardless of what these conditions
might be.

By convention we choose to denote the sublattice contributions as
follows:
\begin{itemize}
\item $f_{m,n}$ for an $m\times n$ lattice with free boundary conditions;
\item $f^{\Left}_{m,n}$ (resp.\ $f^{\Low}_{m,n}$) for a particular boundary
 condition on one vertical (resp.\ horizontal) side;
\item $f^{\LeftLow}_{m,n}$ for particular boundary conditions on two adjacent
 sides (i.e., one vertical and one horizontal);
\item $f^{\LeftRight}_{m,n}$ (resp.\ $f^{\LowUp}_{m,n}$) for particular
 boundary conditions on two opposite vertical (resp.\ horizontal) sides;
\item $f^{\LeftLowRight}_{m,n}$ (resp.\ $f^{\LeftLowUp}_{m,n}$) for particular
 boundary conditions on three adjacent sides;
\item $f^{\All}_{m,n}$ for particular boundary conditions on all four sides.
\end{itemize}
Similar notations apply to the $\tilde{f}$ contributions.

Just as in the case of free boundary conditions, we start in the case
of a bounded $M\times N$ infinite lattice by writing the decomposition
\begin{equation}
f^{(\alpha)}_{M,N}=\sum_{i\leq M, j\leq N}{\sum_{\beta}{(\#_{[i,j,\beta],[M,N,\alpha]})\tilde{f}^{(\beta)}_{i,j}}} \,,
\label{fMNbexact}
\end{equation} 
where $\#_{[i,j,\beta],[M,N,\alpha]}$ denotes the number of ways that
an $i\times j$ lattice with $(\beta)$ boundaries can fit into the
considered $M\times N$ lattice with $(\alpha)$ boundaries.
The self-consistent equations relating the finite-lattice
contributions $\tilde{f}^{(\beta)}_{i,j}$ and $f^{(\gamma)}_{m,n}$ are
to be written in the same fashion (see below).

\subsubsection{Particular boundary condition on one side.}

For one particular boundary condition on (say) the left side,
\eref{fMNbexact} takes the form
\begin{eqnarray}
 f^{\Left}_{M,N} &=&
 \sum_{i\leq M-1, j\leq N}{(M-i)(N+1-j)\tilde{f}_{i,j}} \nonumber \\
 &+& \sum_{i\leq M, j\leq N}{(N+1-j)\tilde{f}^{\Left}_{i,j}} \,,
\label{f1}
\end{eqnarray}
where the first (resp.\ second) term corresponds to contributions
where the $[i,j]$ rectangle does not (resp.\ does) touch the left side
of the $[M,N]$ rectangle.

The $\tilde{f}_{i,j}$ were computed in section \ref{sec:free_bcs}. In
the same way, the $\tilde{f}^{\Left}_{i,j}$ are determined
self-consistently from \eref{f1}---with the $M\times N$ lattice being
replaced by an $m\times n$ finite lattice---and are found to be
\begin{equation}
\tilde{f}^{\Left}_{i,j}=\sum_{m\leq i, n\leq j}{\eta(n,i)(\delta_{m,i}-\delta_{m,i-1})(f^{\Left}_{m,n}-f_{m-1,n})} \,.
\label{ft1}
\end{equation} 
Introducing the cutoff set $B(k)$, the free energy of the bounded
infinite lattice is thus approximated as
\begin{eqnarray}
 f^{\Left}_{M,N} &\approx&
 \sum_{[m,n],[i,j]\leq B(k)}{f_{m,n}(M-i)(N+1-j)\eta(m,i)\eta(n,j)}
 \label{f1approx} \\
 &+& \sum_{[m,n],[i,j]\leq B(k)}{(f^{\Left}_{m,n}-f_{m-1,n})(N+1-j)
 (\delta_{m,i}-\delta_{m,i-1})\eta(n,j)} \,. \nonumber
\end{eqnarray} 
After performing the summation over the $[i,j]$ sublattices and having
isolated the free boundary contribution from the corrections related
to the particular boundary, we have
\begin{equation}
f^{\Left}_{M,N} = f_{M,N}+N\delta f_{\rm s}^{\Left}+\delta f_{\rm c}^{\Left} \,,
\end{equation}
where 
\begin{eqnarray}
 \delta f_{\rm s}^{\Left} &\approx&
 \sum_{[m,n]\leq B(k)} (f^{\Left}_{m,n}-f_{m-1,n}) \times
 \nonumber \\
 & & \qquad \qquad (\delta_{m,k-n}-2\delta_{m,k-n-1}+\delta_{m,k-n-2}) - f_{\rm b}
\end{eqnarray}
and
\begin{eqnarray}
 \delta f_{\rm c}^{\Left} &\approx&
 \sum_{[m,n]\leq B(k)} (f^{\Left}_{m,n}-f_{m-1,n})(m-k+1) \times
 \nonumber \\
 & & \qquad \qquad 
 (\delta_{m,k-n}-2\delta_{m,k-n-1}+\delta_{m,k-n-2}) -f_{\rm s}  \,.
\end{eqnarray}
Notice that the bulk energy is left unchanged by the introduction of
the boundary, and that the surface correction is logically
proportional to the length of the corresponding side. Similar remarks
will hold with two or more particular boundaries.
 
\subsubsection{Particular boundary condition on two adjacent sides.}

In the same way, we have for two adjacent particular boundaries
\begin{eqnarray}
 f^{\LeftLow}_{M,N} &=&
 \sum_{i\leq M-1, j\leq N-1}{(M-i)(N-j)\tilde{f}_{i,j}} +
 \sum_{i\leq M, j\leq N}{\tilde{f}^{\LeftLow}_{i,j}} \nonumber \\
 &+& \sum_{i\leq M, j\leq N-1}{(N-j)\tilde{f}^{\Left}_{i,j}} +
 \sum_{i\leq M-1, j\leq N}{(M-i)\tilde{f}^{\Low}_{i,j}} \,,
\label{f2}
\end{eqnarray} 
with the $\tilde{f}_{i,j}$ and $\tilde{f}^{\Left}_{i,j}$ already
detemined above. Self-consistency of eq.~\eref{f2} for finite
lattices requires that
\begin{eqnarray}
 \tilde{f}^{\LeftLow}_{i,j} &=&
 \sum_{m\leq i, n\leq j} (\delta_{m,i}-\delta_{m,i-1})
 (\delta_{n,j}-\delta_{n,j-1}) \times
 \nonumber \\
 & & \qquad \quad
 (f^{\LeftLow}_{m,n}+f_{m-1,n-1}-f^{\Left}_{m,n-1}-f^{\Low}_{m-1,n}) \,.
\label{ft2}
\end{eqnarray}
Finally, 
\begin{equation}
 f^{\LeftLow}_{M,N} = 
 f_{M,N}+(M+N)\delta f_{\rm s}^{\Left}+\delta f_{\rm c}^{\LeftLow} \,,
\end{equation}
where 
\begin{eqnarray}
 \delta f_{\rm c}^{\LeftLow} &=&
 \sum_{[m,n]\leq B(k)} (f^{\LeftLow}_{m,n}+f_{m-1,n-1}-
 f^{\Left}_{m,n-1}-f^{\Low}_{m-1,n}) \times
 \nonumber \\
 & & \qquad \quad \quad
 (\delta_{m,k-n}-\delta_{m,k-n-1}) +
 2\delta f_{\rm c}^{\Left}-2\delta f_{\rm s}^{\Left}-f_{\rm b} \,.
\end{eqnarray}

\subsubsection{Particular boundary condition on two opposite sides.}

For particular boundary conditions on two opposite sides (say, the right and
left sides), similar calculations lead to
\begin{equation}
 f^{\LeftRight}_{M,N} =
 f_{M,N}+2N\delta f_{\rm s}^{\Left} + \delta f_{\rm c}^{\LeftRight} \,,
\end{equation}
where 
\begin{equation}
 \delta f_{\rm c}^{\LeftRight} = 2\delta f_{\rm c}^{\Left} \,.
\end{equation}

\subsubsection{Particular boundary condition on three sides.}

For particular boundary conditions on three (say, the right, lower and
left sides),
\begin{equation}
 f^{\LeftLowRight}_{M,N} =
 f_{M,N}+(2N+M)\delta f_{\rm s}^{\Left}+\delta f_{\rm c}^{\LeftLowRight} \,,
\end{equation}
where 
\begin{equation}
 \delta f_{\rm c}^{\LeftLowRight} =
 2\delta f_{\rm c}^{\LeftLow}-\delta f_{\rm c}^{\Left} \,.
\end{equation}

\subsubsection{Particular boundary condition on all four sides.}

Finally, for particular boundary conditions on all four sides, 
\begin{equation}
 f^{\All}_{M,N} = 
 f_{M,N}+(2N+2M)\delta f_{\rm s}^{\Left}+\delta f_{\rm c}^{\All} \,,
\end{equation}
where 
\begin{equation}
 \delta f_{\rm c}^{\All} =
 4\delta f_{\rm c}^{\LeftLow}-4\delta f_{\rm c}^{\Left} \,.
\end{equation}

\subsection{General observations}
\label{conclusion enting boundaries}

The remarks made in the case of $f^{\Left}_{M,N}$ can be extended to
all the boundary configurations. In general, the above calculations
show that:
\begin{itemize}

\item The bulk free energy is not affected by any change of boundary
  conditions.

\item The surface energy can be written as a sum over independent
  sides, the energies associated with each type of side being
  distributed according to
\begin{eqnarray}
 f_{\rm s}^{\freeside} &=& f_{\rm s}/2 \nonumber \\
 f_{\rm s}^{\bdside} &=&
 f_{\rm s}^{\freeside} + \delta f_{\rm s}^{\bdside} \,, \qquad
 \mbox{with } \delta f_{\rm s}^{\bdside} \equiv \delta f_{\rm s}^{\Left}  \,. 
\end{eqnarray}

\item The corner energy can also be written as a sum over independent corners, the energies associated with each type of corner being distributed according to
\begin{eqnarray}
f_{\rm c}^{\freecorn} = & f_{\rm c}^{\Free}/4  \nonumber \\
f_{\rm c}^{\mixtcorn} = & f_{\rm c}^{\freecorn} + \frac{1}{2}\delta f_{\rm c}^{\Left}  \\
f_{\rm c}^{\bdcorn} = & f_{\rm c}^{\freecorn} + \delta f_{\rm c}^{\LeftLow} - \delta f_{\rm c}^{\Left}  \,.  \nonumber 
\end{eqnarray}

\end{itemize}

The interaction energy between two adjacent sides of the lattice is
taken into account by the corner energy. However, one does not observe
any corner-corner interaction term, which is related to the assumption
implicitly made in the FLM formalism that the $M\times N$ lattice is
bigger than any cutoff set $B(k)$, that is, effectively infinite.

The absence of corner-corner interactions is radically different from
the outcome of CFT calculations, where the partition function on a
large rectangle depends non-trivially on its aspect ratio through the
so-called modular parameter. This implies in particular (but not only)
that the change of boundary conditions at the corners interact, and
in fact define a correlation function of boundary condition changing
operators. In view of this fundamental difference, it is all the more
remarkable that the asymptotics of the corner free energies reported
here link up nicely with CFT predictions \cite{CardyPeschel}.

\subsection{Case of the Ising model}
\label{FLMboundIsing}

As previously announced, the case of the Ising model with fixed boundary
conditions is not correctly described by the above calculations.

To get a feeling of what goes wrong, we first consider the example
``$+-$ff'', where the two adjacent sides of the $M \times N$ rectangle
support respectively $+$ and $-$ fixed boundary conditions, and the
remaining two sides are free. This implies that a domain wall will
originate from the $+-$ corner and terminate on any of the two free
sides.  In particular, its length will be at least ${\rm
  min}(M,N)$. This situation is at odds with the perturbative
principle illustrated in Fig.~\ref{ground_state}, where the
excitations are with respect to a trivial ground state.

So contrary to what was done in the case of loop models,
finite-sublattice excitations in the Ising model with one or more sides
supporting fixed boundary conditions cannot be thought of as localised
perturbations independent of the surrounding ground state.  Instead,
these excitations interact with the surrounding spins, therefore
affecting their configuration of minimal energy.  We leave the
resolution of this problem for future work.

In the following, we shall therefore focus on boundary conditions in
loop models, where excitation can still be described as local
perturbations of the ground state in Fig.~\ref{ground_state}. This
leads us naturally to consider the JS boundary conditions
\cite{JacobsenSaleurConformalBoundary} for the selfdual Potts model
on the square lattice.

\subsection{JS boundary conditions}
\label{Parametrization}

Within rational CFT the number of possible conformal boundary
conditions is known to be equal to the number of primary operators.
This result does however not apply to
the $Q$-state Potts model for generic values of $Q$. Instead one expects
the existence of infinitely many distinct conformal boundary conditions.

One infinite family of such conformally invariant boundary conditions
was given in \cite{JacobsenSaleurConformalBoundary}. Let us
parameterise the bulk loop weight $n$ by
\begin{equation} 
 n = 2\cos \left( \frac{\pi}{p+1} \right) \,,
\label{conformal n}
\end{equation}
so that $p \in \mathbb{R}$ reduces to the usual minimal model index
whenever $p \in \mathbb{N}$. The JS boundary conditions then amount to
assigning a different weight $n_1$ to each loop touching at least once
any one of the particular edges.  Following
\cite{JacobsenSaleurConformalBoundary} we parameterise $n_1$ by
\begin{equation} 
 n_1 = \frac{\sin \left( \frac{(r+1)\pi}{p+1} \right)}
            {\sin \left( \frac{r\pi}{p+1} \right)} \,,
\label{conformal n1}
\end{equation}
with $0 \le r < p+1$.

In the following we shall only be concerned with the case of $r \in
\mathbb{N}$.  The case $r=1$ corresponds to $n_1=n$, or Dirichlet
boundary conditions, and the corrections to surface and corner free
energies should therefore be exactly zero. When $r = p/2$ we have
$n_1=1$, which corresponds to Neumann boundary conditions. Note also
the case $r=0$ which corresponds to $n_1 \to \infty$. Each spin on a
particular boundary is then enclosed by a {\rm distinct} boundary
loop, a problem which seems likely to have physical applications.

The parameterisations (\ref{conformal n})--(\ref{conformal n1}) need
to be linked up with the fact that the FLM method permits us to approach
the conformal limit $n \to 2^+$ from the side $n>2$ (i.e., $Q>4$).  We
therefore set $\gamma \equiv \frac{\pi}{p+1}$ and let $\gamma =
\mathrm{i} \widetilde{\gamma}$, with $\widetilde{\gamma}\in
\mathbb{R}_+$. Thus
\begin{eqnarray}
 n &=& 2 \cosh \widetilde{\gamma} \,,
 \nonumber \\
 n_1 &=& \frac{\sinh((r+1)\widetilde{\gamma})}{\sinh(r \widetilde{\gamma})} \,.
 \label{param n}
\end{eqnarray}
In the conformal limit $\widetilde{\gamma} \to 0^+$ we thus have $n_1
\sim \frac{r+1}{r}$, meaning in particular that $n_1 > 1$ for any
finite $r \in \mathbb{N}$, whereas the Neumann boundary condition
$n_1 = 1$ is formally obtained in the limit $r \to \infty$. These
remarks will turn out important for the following discussion.

\subsection{Results for JS boundary conditions}
\label{Resultsbound}

In this section we report our results for the corrections to the surface and corner free
energies for the selfdual square-lattice Potts model with
JS boundary conditions.  We have performed the explicit FLM calculations
for $r=0,1,\ldots,9$ and from those we have been able to conjecture
product formulae that we believe are valid for any $r \in
\mathbb{N}$.

\subsubsection{Bulk free energy.}

The bulk free energy is invariably found to be unchanged upon taking
$r \neq 1$ in the JS boundary condition.  This was of course to be
expected on physical grounds, and indeed follows analytically from the
results of section \ref{conclusion enting boundaries}.

\subsubsection{Surface free energy.}

The factorised form of the correction to the surface free energy
(i.e., $\delta f_{\rm s}^{\bdside} = f_{\rm s}^{\bdside} - f_{\rm
  s}^{\freeside}$ with $f_{\rm s}^{\bdside} = f_{\rm s}^{\rm JS}$ and
$f_{\rm s}^{\freeside} = f_{\rm s}^{\rm free}$) for the selfdual Potts
model on the square lattice reads, as a function of $r$:
\begin{itemize}
\item for $r=0$,  
\begin{equation}
\rme^{\delta f_{\rm s}^{\bdside}} = \frac{1-q}{1-q^4} \prod_{k=1}^{\infty}{\left(\frac{(1-q^{4k+1})(1-q^{4k+2})}{(1-q^{4k-1})(1-q^{4k+4})}\right)^2} \,,
\end{equation}

\item for $r=1$,
\begin{equation}
\rme^{\delta f_{\rm s}^{\bdside}} = 1 \,,
\end{equation}

\item for $r$ an even positive integer, $r=2p$,
\begin{eqnarray}
\rme^{\delta f_{\rm s}^{\bdside}} = &  \left( \prod_{k=1}^{p-1}{\frac{(1-q^{4k-1})(1-q^{4k+4})}{(1-q^{4k+1})(1-q^{4k+2})}} \right) \frac{1-q^{4p-1}}{1-q^{4p+2}} \times \nonumber \\  & \prod_{k=1}^{\infty}{\left(\frac{(1-q^{4k+1})(1-q^{4k+2})}{(1-q^{4k-3})(1-q^{4k+4})}\right)^2} \,, 
\label{reven}
\end{eqnarray}

\item for $r$ an odd positive integer different from $1$, $r=2p+1$,
\begin{equation}
\rme^{\delta f_{\rm s}^{\bdside}} =  \prod_{k=1}^{p}{\frac{(1-q^{4k+1})(1-q^{4k+2})}{(1-q^{4k-1})(1-q^{4k+4})}} \,.
\label{rodd}
\end{equation}

\end{itemize}

When $r\to \infty$, \eref{reven} and \eref{rodd} give the same limit, namely 
\begin{equation}
 {\rm e}^{\delta f_{\rm s}^{\bdside}} = \prod_{k=1}^{\infty}
 {\frac{(1-q^{4k+1})(1-q^{4k+2})}{(1-q^{4k-1})(1-q^{4k+4})}} \,.
\end{equation}

\subsubsection{Corner free energy.}

In the same way we find the corrections to the corner free energy when
one side supports the JS boundary condition:

\begin{itemize}
\item for $r$ an even positive integer, $r=2p$,
\begin{eqnarray}
\rme^{\delta f_{\rm c}^{\Left}} &=& \prod_{k=2p}^{4p}{\frac{1}{1-q^{2k+1}}} \prod_{k=p+1}^{\infty}{\frac{1}{(1-q^{8k-1})(1-q^{8k+1})}} \nonumber \\  & & \prod_{k=1}^{\infty}{(1-q^{8k-5})(1-q^{8k-3})} \,,
\label{efc1_even}
\end{eqnarray}

\item for $r$ an odd positive integer, $r=2p+1$,
\begin{equation}
\rme^{\delta f_{\rm c}^{\Left}} =  \prod_{k=2p+1}^{4p+2}{\frac{1}{1-q^{2k+1}}}  \prod_{k=1}^{p+1}{(1-q^{8k-5})(1-q^{8k-3})} \,,
\label{efc1_odd}
\end{equation}
which includes the special case $\rme^{\delta f_{\rm c}^{\Left}}=1$ for $r=1$.

\end{itemize}

When $r\to \infty$, the common limit of (\ref{efc1_even})--(\ref{efc1_odd}) is
\begin{equation}
\rme^{\delta f_{\rm c}^{\Left}} =  \prod_{k=1}^{\infty}{(1-q^{8k-5})(1-q^{8k-3})} \,.
\label{cornermixtrinf}
\end{equation}
Similarly, when JS boundary conditions are imposed on two adjacent sides:

\begin{itemize}

\item for $r$ an even positive integer, $r=2p$,
\begin{eqnarray}
 \rme^{\delta f_{\rm c}^{\LeftLow}} &=& \frac{1-q^{4p+2}}{1-q^4} \times
 \nonumber \\
 & & \prod_{k=p+1}^{\infty}{\frac{(1-q^{8k-5})^2(1-q^{8k-3})^2(1-q^{4k})^2}{(1-q^{4k-1})^2 (1-q^{4k-3})^2(1-q^{8k-6})(1-q^{8k-2})}}  \times \nonumber \\
 & & \prod_{k=1}^{\infty}{\frac{(1-q^{8k-5})^2(1-q^{8k-3})^2}{(1-q^{8k-2})(1-q^{8k+2})}} \,, 
 \label{efc2_even}
\end{eqnarray}

\item for $r$ an odd positive integer, $r=2p+1$,
\begin{eqnarray}
\rme^{\delta f_{\rm c}^{\LeftLow}} &=& \frac{1-q^{4p+4}}{1-q^4} \prod_{k=1}^{p}{\frac{(1-q^{4k-1})^2(1-q^{4k+1})^2}{(1-q^{8k-1})^2(1-q^{8k+1})^2}}  \times \nonumber \\
& & \prod_{k=1}^{p}{\frac{(1-q^{8k-2})(1-q^{8k+2})}{(1-q^{4k+2})^2}} \,,
 \label{efc2_odd}
\end{eqnarray}
which includes the special case $\rme^{\delta f_{\rm c}^{\LeftLow}}=1$ for $r=1$.
\end{itemize}

When $r\to \infty$, the common limit of (\ref{efc2_even})--(\ref{efc2_odd}) is
\begin{equation}
\rme^{\delta f_{\rm c}^{\LeftLow}} =  \frac{1}{1-q^4} \prod_{k=1}^{\infty}{\frac{(1-q^{8k-5})^2(1-q^{8k-3})^2}{(1-q^{8k-2})(1-q^{8k+2})}} \,.
\label{cornerbdrinf}
\end{equation}

\subsection{Critical limit} \label{Criticalbound}

In the critical limit, $q \to 1^-$, we can now extract the finite
limits of the surface free energy corrections, and the asymptotic
divergent behaviour of the corner free energy corrections.

\subsubsection{Finite limits of surface free energy corrections.}

The $q\to 1^{-}$ limit of $\rme^{\delta f_{\rm s}^{\bdside}}$ is found to be:

\begin{itemize}

\item for $r=0$,  
\begin{equation}
\lim_{q\to 1^{-}} \rme^{\delta f_{\rm s}^{\bdside}} = \frac{64}{\pi}\frac{\Gamma\left(\frac{3}{4}\right)^2}{\Gamma\left(\frac{1}{4}\right)^2} \,,
\end{equation}

\item for $r$ an even positive integer, $r=2p$,
\begin{equation}
\lim_{q\to 1^{-}} \rme^{\delta f_{\rm s}^{\bdside}} = 8\sqrt{2\pi}\frac{\Gamma\left(\frac{3}{4}+p\right) \Gamma\left(1+p\right)}{\Gamma\left(\frac{1}{4}\right)^2\Gamma\left(\frac{1}{4}+p\right)\Gamma\left(\frac{3}{4}+p\right)} \,,
\end{equation}

\item for $r$ an odd positive integer, $r=2p+1$,
\begin{equation}
\lim_{q\to 1^{-}} \rme^{\delta f_{\rm s}^{\bdside}} = \frac{2}{\sqrt{\pi}}\frac{\Gamma\left(\frac{3}{4}\right)\Gamma\left(\frac{5}{4}+p\right) \Gamma\left(\frac{3}{2}+p\right)}{\Gamma\left(\frac{5}{4}\right)\Gamma\left(\frac{3}{4}+p\right)\Gamma\left(2+p\right)}  \,,
\end{equation}
which includes the special case $\lim_{q\to 1^{-}} \rme^{\delta f_{\rm c}^{\bdside}} =1$ for r=1.

\end{itemize}

When $r$ is taken to infinity,
\begin{equation}
\lim_{q\to 1^{-}} \rme^{\delta f_{\rm s}^{\bdside}} =  \frac{8}{\sqrt{\pi}} \frac{\Gamma\left(\frac{3}{4}\right)}{\Gamma\left(\frac{1}{4}\right)} \,.
\end{equation}

\subsubsection{Divergence of the corner free energy corrections.}
\label{sec:div_efc_corr}

More interesting is the effect of JS boundary conditions on the
critical divergence of the corner free energy, which gives access to
the effect of these particular boundary conditions on the effective
central charge.

In the case of a corner between a free edge and one edge with JS
boundary conditions, the correction to $\rme^{f_{\rm c}}$ is found to
have a {\em finite limit}
\begin{equation}
\lim_{q\to 1^{-}} \rme^{\delta f_{\rm c}^{\Left}} = \frac{4^{1+r}\sqrt{2+\sqrt{2}}}{\pi}\frac{\Gamma\left(\frac{9+4r}{8}\right) \Gamma\left(\frac{7+4r}{8}\right)\Gamma\left(\frac{3+2r}{2}\right)}{\Gamma\left(\frac{3+4r}{2}\right)} \,.
\label{eq:divergence mixtcorn r}
\end{equation}
for any {\em finite} value of $r$ (whether even or odd).  This finite
limit just adds a constant term to the corner free energy, with no
effect on the diverging part.

On the contrary, when $r$ is taken to infinity, we find from
\eref{cornermixtrinf} that
\begin{equation}
 \rme^{\delta f_{\rm c}^{\Left}}  \stackrel[q\to 1^{-}]{}{\sim}
 \sqrt{1-\frac{1}{\sqrt{2}}}
 \rme^{-\frac{\pi^2}{24}\left(\frac{1}{1-q}-\frac{1}{2}\right)} \,,
 \label{cdfc_one_JS_side}
\end{equation}
so that the diverging part is modified. We shall discuss these findings
more fully below.

In the case of a corner between two edges with JS boundary
conditions we find for finite $r$
\begin{equation}
\lim_{q\to 1^{-}} \rme^{\delta f_{\rm c}^{\LeftLow}} = \frac{r+1}{2} 2^{-4r}\sqrt{2+\sqrt{2}} \frac{ \pi^{3/2} \Gamma\left(\frac{1+2r}{2}\right)^3}{ \Gamma\left(\frac{3+4r}{8}\right)^2\Gamma\left(\frac{5+4r}{8}\right)^2\Gamma\left(\frac{2+r}{2}\right)^2} \,.
\end{equation}
On the contrary, when $r$ is taken to infinity, we find from
\eref{cornerbdrinf} that
\begin{equation}
 \rme^{\delta f_{\rm c}^{\LeftLow}} \stackrel[q\to 1^{-}]{}{\sim}
 \frac{\sqrt{2}-1}{4}  \rme^{-\frac{\pi^2}{24}\left(\frac{1}{1-q}-\frac{1}{2}\right)}
 \,.
\end{equation}
Note that, apart from the prefactor, the diverging part is {\em identical}
to that of (\ref{cdfc_one_JS_side}).

In conclusion, the diverging parts of the corner free energies
associated with each possible kind of corner are
\begin{eqnarray}
 f_{\rm c}^{\freecorn} & \stackrel[q\to 1^{-}]{}{\sim} &
 \frac{\pi^2}{32(1-q)} \,, \nonumber \\
 f_{\rm c}^{\mixtcorn} & \stackrel[q\to 1^{-}]{}{\sim} &
 \frac{\pi^2}{96(1-q)} \qquad \mbox{for $r \to \infty$} \,,
 \label{eq:corrections r infinite} \\
 f_{\rm c}^{\bdcorn} & \stackrel[q\to 1^{-}]{}{\sim} &
 \frac{\pi^2}{32(1-q)} \qquad \mbox{for $r \to \infty$} \,.
 \nonumber
\end{eqnarray}

\subsection{Relation to conformal field theory.}

We can now compare the divergence of the corner free energies with CFT
results for the JS boundary conditions
\cite{JacobsenSaleurConformalBoundary}.

It was found in \cite{JacobsenSaleurConformalBoundary} that the operator
that changes the boundary conditions from free to JS with parameter $r$
has the conformal weight
\begin{equation}
 h \equiv h_{r,r} = \frac{r^2-1}{4p(p+1)} \,.
\end{equation}
In the limit $n \to 2^-$ (i.e., $p \to \infty$) any fixed boundary
loop weight $n_1 > 1$ corresponds to a value of $r$ that grows slower
than $p$. In other words, $h \to 0$ as $p \to \infty$. On the other
hand, $n_1 = 1$ corresponds to $r = p/2$, in which case we obtain
the finite limit $h \to \frac{1}{16}$ as $p \to \infty$. So summarising,
$h(n_1)$ tends to a step function:
\begin{equation}
 \lim_{p \to \infty} h =
 \left \lbrace
 \begin{array}{ll}
 \frac{1}{4}  & \mbox{for } n_1 < 1 \,, \\
 \frac{1}{16} & \mbox{for } n_1 = 1 \,, \\
 0            & \mbox{for } n_1 > 1 \,. \\
 \end{array}
 \right.
 \label{h_step_function}
\end{equation}

On the other hand, the insertion of a boundary condition changing
operator in a corner will change the effective central charge
according to \cite{Bondesan}
\begin{equation}
 c_{\rm eff} = c - 32 h \,.
 \label{chg_c_eff}
\end{equation}
So according to (\ref{eq:cardy}) we expect a change in the divergence
of the corner free energy if and only if $h \neq 0$.

Recall meanwhile that $n_1 \sim \frac{r+1}{r}$ from (\ref{param n}).
So whenever $r$ is {\em finite} we have $n_1 > 1$, and so by
(\ref{h_step_function}) the divergent part of the corner free energy
should remain unchanged. This is precisely what we have found to be
the case in section \ref{sec:div_efc_corr}.

On the other hand, in the $r \to \infty$ limit we have $n_1 = 1$, and
so from (\ref{h_step_function})--(\ref{chg_c_eff}) the divergent parts
of $f_{\rm c}^{\mixtcorn}$ and $f_{\rm c}^{\freecorn}$ should be
different. While such a difference is indeed apparent in
eq.~(\ref{eq:corrections r infinite}), the actual values do not quite
work out as expected: we have $c=1$, and from (\ref{eq:corrections r
  infinite}) we find $c_{\rm eff}=\frac{1}{3}$, so that $h =
\frac{1}{48}$.  This is at odds with $h=\frac{1}{16}$ given in
(\ref{h_step_function}).

This discrepancy is maybe not completely surprising. Indeed, two very
different double limits are at work in the conformal case ($p \to \infty$ and
$n_1 \to 1$) and in the asymptotic analysis of the corner free energy
($q \to 1^-$ and $r \to \infty$). Meanwhile, the critical exponent tends
to a step function (\ref{h_step_function}), so the only disagreement
concerns the value of $h$ right at the step. It is conceivable that a
non-commutativity of limits misses this value (by a factor of $3$).

It remains to discuss the case of $f_{\rm c}^{\bdcorn}$. Since in this
case there is no insertion of a boundary condition changing operator
in the corner, the CFT prediction (\ref{chg_c_eff}) is that $f_{\rm
  c}^{\bdcorn}$ should have the {\em same} divergence as $f_{\rm
  c}^{\freecorn}$. This is precisely what we have found in
(\ref{eq:corrections r infinite}). On the other hand, it is clear that
the two types of corners {\em are} different, and this should be
reflected by a finite (non-diverging) difference in the two corner
free energies. Once again, this is exactly what we found in
section~\ref{sec:div_efc_corr}.

\section{Conclusion}

In this paper we have presented what is---to our knowledge---the first
systematic study of corner free energies from the perspective of
exactly solvable models. Combining the FLM method with exact
enumeration results, we have obtained exact (albeit conjectured)
product formulae for the corner free energy ${\rm e}^{f_{\rm c}}$ for
several integrable cases of two-dimensional Potts and Ising models,
as well as for the FPL${}^2$ loop model.

We have obtained the asymptotic expansions of these expressions for
${\rm e}^{f_{\rm c}}$ near several conformally invariant critical
points. This has permitted us to identify the asymptotic divergence of
the correlation length, in agreement with Bethe Ansatz results
whenever the latter are available, and provided new results in other
cases. More importantly, the comparison between results for the square
and triangular lattices gave agreement with ideas of universality and
has enabled us to verify the angular dependence of the Cardy-Peschel
formula (\ref{eq:cardy}). Such agreement was found in particular for
the ferromagnetic and antiferromagnetic transitions of the Potts
model, and for the Ising model.

In some cases we have also provided new results for the surface free
energy ${\rm e}^{f_{\rm s}}$.

In the first part of the paper we were concerned with free boundary
conditions.  But in section~\ref{sec:boundaries} we have shown how to
generalise the FLM formalism to the case of special boundary
conditions. We have used this to study in detail the so-called JS
boundary conditions \cite{JacobsenSaleurConformalBoundary}, taking the
selfdual square-lattice Potts model as an example. In particular we
found some agreement with CFT predictions for the case where the
corner contains a boundary condition changing operator.

We leave several issues for future work.  On the side of exactly
solvable models, it would be interesting to establish if the corner
free energy can be obtained exactly from the corner transfer
matrix. If this is possible, one could hope to prove our formulae for
${\rm e}^{f_{\rm c}}$.  On the conformal side, we believe that our
results can be interpreted in terms of boundary states, generalising
the ideas of \cite{Bondesan}.

\subsection*{Acknowledgments}

We thank R.\ Bondesan, J.L.\ Cardy, H.\ Saleur and A.D.\ Sokal for
stimulating discussions.  This work was supported by a grant from the
Agence Nationale de la Recherche (Projet 2010 Blanc SIMI 4: DIME).

\appendix

\section{Factorised form of the bulk free energies
from analytical results}
\label{appendix comparison products analytical}

In this appendix, we describe how certain factorised expressions found
for the bulk free energy of Potts and Ising models can be recovered
from existing exact results. The first step is to rewrite into a
slightly different form the logarithm of generic factorised
expressions of the form
\begin{equation}
\prod_{k=1}^{\infty}(1-q^{\beta k + \gamma})^{\mu k + \nu} \,.
\end{equation}
Using a series expansion, we have 
\begin{eqnarray}
 \log \prod_{k=1}^{\infty}(1-q^{\beta k + \gamma})^{\mu k + \nu}
 &=& \sum_{k=1}^{\infty}{(\mu k + \nu)\log \left(1-q^{\beta k + \gamma}\right) }
 \nonumber \\
 &=& -\sum_{n=1}^{\infty}{\frac{1}{n} \sum_{k=1}^{\infty}
 {(\mu k + \nu) q^{(\beta k + \gamma)n}}}
 \nonumber \\
 &=& -\sum_{n=1}^{\infty}{\frac{q^{\gamma n}}{n}
 \frac{q^{\beta n}(\mu+\nu-\nu q^{\beta n})}{(1-q^{\beta n})^2}} \,,
 \label{identity log product}
\end{eqnarray}
where between the last two lines we have used the exponential
polynomial summation formula to perform the sum over $k$.
From this formula, we can identify some of the bulk free energies
obtained in the main text with existing exact results.

\subsection{Selfdual square-lattice Potts model}
\label{appendix_A1}

Applying the above identity (\ref{identity log product}) to the
product form (\ref{products selfdualsquare}) obtained for the bulk
free energy of the selfdual square-lattice Potts model yields
\begin{eqnarray}
f_{\rm b} &=& 2\ln q + \ln (1+q^2) + 2\sum_{k=1}^{\infty}{\frac{q^k}{k}} - 4\sum_{k=1}^{\infty}{\frac{1}{k}\frac{q^{3k}}{1+q^{2k}}} \nonumber \\
&=& 2\ln q + \ln (1+q^2) + 2\sum_{k=1}^{\infty}{\frac{q^k}{k}\frac{1-q^{2k}}{1+q^{2k}}} \,.
\label{fb pottsselfdualsquare}
\end{eqnarray}

In \cite{BaxterBook}, the dimensionless bulk free energy (that is, the
opposite of our $f_{\rm b}$) of the square-lattice Potts model is
shown to be (\ref{BaxterPottsSquareBulk}), which we reproduce here for
convenience,
\begin{equation}
\psi = -\frac{1}{2} \ln Q - 2\left[\beta + \sum_{k=1}^{\infty}{\frac{1}{k}\rme^{-k\lambda} \frac{\sinh 2 k \beta}{\cosh  k \lambda} } \right] \,,
%\label{BaxterPottsSquareBulk}
\end{equation}
where $\beta$ and $\lambda$ are defined by
\begin{eqnarray}
Q^{1/2} =& 2\cosh\lambda \\
\frac{v}{Q^{1/2}} =& \frac{\sinh\beta}{\sinh(\beta-\lambda)} \,.
\end{eqnarray}
On the selfdual curve, $v=\sqrt{Q}$, that is, $\beta =
\frac{\lambda}{2}$ with $-\lambda=\ln q$. We thus have
\begin{equation}
\frac{\sinh 2 k \beta}{\cosh  k \lambda} = \tanh k\lambda = \frac{1-q^{2k}}{1+q^{2k}} \,,
\end{equation}
and thus
\begin{equation}
\psi = -\ln\left(q+\frac{1}{q}\right)-2\left[-\frac{1}{2}\ln q + \sum_{k=1}^{\infty}{\frac{q^k}{k}\frac{1-q^{2k}}{1+q^{2k}}}\right] \,,
\end{equation}
that is $\psi = -f_{\rm b}$ as claimed.

\subsection{Antiferromagnetic square-lattice Potts model}
\label{appendix_A2}

For the critical antiferromagnetic Potts model, our aim is to approach
the critical point $(Q,v) = (0,0)$ from the left side of the
$v=-2+\sqrt{4-Q}$ curve, that is with $Q<0$, using the parameterisation
\begin{equation}
 Q = -\left(q-q^{-1}\right)^2 \,.
\end{equation} 
Assuming that $f_{\rm b}$ has no singularity at $Q = \infty$,
this region can be mapped to the $Q>4$ region by setting 
\begin{equation}
q = \mathrm{i}p \,,
\end{equation} 
and taking $p$ real (more precisely, $p\in [0,1[$). Since $|q|<1$ all
along the problem, the FLM expansions are kept convergent and should
remain valid. In terms of this parameterisation, we have on the
$v=-2+\sqrt{4-Q}$ curve
\begin{eqnarray} 
 Q^{1/2} &=& p+p^{-1}, \nonumber \\
 v &=&  -2 + \mathrm{i}(p-p^{-1})\,. 
\label{paramAFp}
\end{eqnarray}
We now can use Baxter's formula in \cite{AFBaxter}, namely 
\begin{equation}
 \psi = -\frac{1}{2} \ln Q - 2\left[u + \sum_{k=1}^{\infty}{\frac{1}{k}\rme^{-k\lambda} \frac{\sinh 2 k u}{\cosh  k \lambda} } \right] \,,
\label{BaxterPottsAFBulk}
\end{equation}
where
\begin{eqnarray}
 \psi &=& -f_{\rm b} \,, \nonumber \\
 Q^{1/2} &=& 2\cosh\lambda \,,  \\
 \rme^K &=& \frac{\sinh(\lambda+u)}{\sinh(\lambda-u)} \,. \nonumber
\label{definBaxterAF}
\end{eqnarray}
In terms of $p$, we find from (\ref{paramAFp}) and
(\ref{definBaxterAF}) the following expression for $u$,
\begin{eqnarray}
u = \frac{3\mathrm{i}\pi}{4}-\frac{1}{2}\ln p = \frac{3\mathrm{i}\pi}{4}+\frac{\lambda}{2} \,,
\end{eqnarray}
which indeed satisfies the conditions $0 < \Re u < \lambda$ and $0\leq
\Im u < \pi$. Eq.~(\ref{BaxterPottsAFBulk}) can then be rewritten
\begin{eqnarray}
 \psi &=& -\frac{1}{2} \ln Q - 2\left[\frac{3\mathrm{i}\pi}{4}-\frac{1}{2}\ln p + \sum_{k=1}^{\infty}{\frac{(-\mathrm{i}p)^k}{k}\frac{1-(-1)^k p^{2k}}{1+p^{2k}} } \right]  \nonumber \\
&=& -\ln (1+p^2)+2\ln p -\frac{3\mathrm{i}\pi}{2} -2\sum_{k=1}^{\infty}{\frac{(-\mathrm{i}p)^k}{k}\frac{1-(-1)^k p^{2k}}{1+p^{2k}} } \nonumber \\
&=& -\ln (1-q^2)+ \ln q^2 -\frac{\mathrm{i}\pi}{2} -2\sum_{k=1}^{\infty}{\frac{q^k}{k}\frac{(-1)^k-q^{2k}}{1+q^{2k}} } \,.
\label{rewriteAFBaxter}
\end{eqnarray}
Writing the sum on the right-hand side as
\begin{eqnarray}
\sum_{k=1}^{\infty}{\frac{q^k}{k}\frac{(-1)^k-q^{2k}}{1+q^{2k}} } &=&  \sum_{k=1}^{\infty}{\frac{(-q)^k}{k}\frac{(1-q^{2k})(1-(-1)^k q^{2k})}{1-q^{4k}} } \\
&=& \sum_{k=1}^{\infty}{\frac{(-q)^k}{k}\frac{1}{1-q^{4k}}} + \sum_{k=1}^{\infty}{\frac{q^{k}}{k}\frac{q^{4k}}{1-q^{4k}}} \nonumber \\
& &  - \sum_{k=1}^{\infty}{\frac{q^{2k}}{k}\frac{q^{4k}}{1-q^{8k}}}
\end{eqnarray}
we see from (\ref{identity log product}) that it can be put in the form
\begin{equation}
\ln\left[(1-q)\prod_{k=1}^{\infty}{\frac{1-q^{8k-2}}{1-q^{8k-6}}}\right] \,.
\end{equation}
Inserting this back into (\ref{rewriteAFBaxter}), $\rme^{-\psi}$ is
finally shown to be equal to the factorised expression
(\ref{productsAF}) of $\rme^{f_{\rm b}}$, up to a $-\mathrm{i}$ phase
factor which is presumably due to the crossing of some branch cuts
during our analytic continuation.

\subsection{Selfdual triangular-lattice Potts model}
\label{appendix_A3}

For the selfdual Potts model on a triangular lattice, the free energy
in the thermodynamic limit is found in \cite{BaxterBook} to be given
by (\ref{BaxterPottsTri2}), which we reproduce here for
convenience,
\begin{equation}
 \psi = -\frac{1}{2} \ln Q - 3\left[\beta + \sum_{k=1}^{\infty}{\frac{1}{k}\rme^{-k\lambda} \frac{\sinh 2 k \beta}{\cosh  k \lambda} } \right] \,,
%\label{BaxterPottsTri2}
\end{equation}
with the same notations as in section
\ref{sec:selfdual_potts_square}. In terms of $q$, we have on the
selfdual line (\ref{triangularPottscritical})
\begin{eqnarray}
\rme^{-\lambda} &=& q \,, \nonumber \\
\rme^{-\beta} &=& q^{\frac{1}{3}} \,,
\end{eqnarray}
such that eq.~(\ref{BaxterPottsTri2}) reads
\begin{equation}
 \psi = -\frac{1}{2} \ln Q - 3\left[-\frac{1}{3}\ln q +
 \sum_{k=1}^{\infty}
 \frac{q^{2k}}{k}\frac{q^{-2n/3}-q^{2n/3}}{1+q^{2k}} \right] \,.
%\label{BaxterPottsTri2}
\end{equation}
Applying (\ref{identity log product}) to the factorised expression
(\ref{efs_tri_sd}) for $\rme^{f_{\rm b}}$, we check that it agrees with
the above result, that is, $-\psi = f_{\rm b}$.

\subsection{Ising model on the square lattice}
\label{appendix_A4}

Applying eq.~(\ref{identity log product}) to the expression
(\ref{efb_Isingsquare}) found for $\rme^{f_{\rm b}}$ in the
case of the square-lattice Ising model yields
\begin{eqnarray}
f_{\rm b} &=& -\frac{1}{2}\ln q - \sum_{n=1}^{\infty} \frac{1}{n} \left[ \frac{q^{7n}(7+q^{8n})}{(1-q^{8n})^2} + \frac{q^{3n}(3+5q^{8n})}{(1-q^{8n})^2} \right. \nonumber \\ &+& \left. \frac{q^{4n}(2-2q^{8n})}{(1-q^{8n})^2}  
  -\frac{q^{n}(1+7q^{8n})}{(1-q^{8n})^2}  - \frac{q^{5n}(5+3q^{8n})}{(1-q^{8n})^2} - \frac{q^{2n}}{1-q^{8n}} - \frac{q^{6n}}{1-q^{8n}} \right] \nonumber \\ 
 &=& -\frac{1}{2}\ln q + \sum_{n=1}^{\infty} \left[
 \frac{1}{n} \frac{1}{1+q^{4n}} -
 \frac{1+q^n(-1+q^2+q^{2n})}{(1+q^{2n})^2} \right] \,. 
\end{eqnarray}
In the same fashion, Baxter's analytical expression (\ref{fb Ising analytical}) can be transformed as follows
\begin{eqnarray}
-\psi &=&  -\ln x^2 + \sum_{n=1}^{\infty}{\frac{q^{2n}(1-q^{n})^2(1-q^{2n})^2}{n(1+q^{2n})(1-q^{8n})}} \nonumber \\
&=& -\frac{1}{2}\ln q + \ln\prod_{k=1}^{\infty}{\frac{(1-q^{8k-5})(1-q^{8k-3})}{(1-q^{8k-7})(1-q^{8k-1})}} + \sum_{n=1}^{\infty}{\frac{q^{2n}(1-q^{n})^2(1-q^{2n})^2}{n(1+q^{2n})(1-q^{8n})}} \nonumber \\
&=& -\frac{1}{2}\ln q + \sum_{n=1}^{\infty}{\frac{1}{n} \left[ \frac{1}{1+q^{4n}} - \frac{1+q^n(-1+q^2+q^{2n})}{(1+q^{2n})^2} \right]}  \,, 
\end{eqnarray}
which establishes the equivalence of the two expressions. 

\subsection{Ising model on the triangular lattice}
\label{appendix_A5}

Applying (\ref{identity log product}) to the expression
(\ref{product_efb_ising_tri}) found for $\rme^{f_{\rm b}}$ in the case
of the triangular-lattice Ising model yields
\begin{eqnarray}
f_{\rm b} &=& -\frac{3}{2}\ln x - \sum_{n=1}^{\infty}{\frac{q^{8n}}{n(1-q^{8n})}\left[2q^{-4n}-q^{-6n}-q^{-2n}\right]}  \nonumber\\
 &-& \sum_{n=1}^{\infty}\frac{q^{8n}}{n(1-q^{8n})^2} \left[\left(6-3(1-q^{8n})(q^{-14n/3}-q^{-10n/3})\right) \right. \nonumber \\
&+& \left[ 6(q^{-4n/3}+q^{-2n/3}+q^{8n/3}-q^{-8n/3}-q^{2n/3}-q^{4n/3})\right] \nonumber \\
&=& -\frac{3}{2}\ln q -\sum_{n=1}^{\infty}{\frac{q^{2n}}{n}\frac{(1-q^{2n})^2}{1-q^{8n}}} \nonumber \\
&-&  \sum_{n=1}^{\infty}{\frac{3 q^{3n}}{n(1-q^{8n})}\frac{1-q^{2n}}{1+q^{2n}}}  \,,
\end{eqnarray}
which is seen to be equivalent to the analytical expression
(\ref{analytical_efb_ising_tri}) derived by Baxter.

\end{document}